\documentclass[numberedappendix]{emulateapj}
\usepackage[colorlinks,citecolor=cyan]{hyperref}

\usepackage{graphicx}
\usepackage{amsmath}
\usepackage{psfig}
\usepackage{apjfonts}
\usepackage{color}

\def\sec{\textrm{s}}

\def\nm{\textrm{nm}}
\def\um{\mu\textrm{m}}

\def\cm{\textrm{cm}}
\def\meter{\textrm{m}}
\def\km{\textrm{km}}

\def\kms{\textrm{km\ s}^{-1}}

\def\Lsun{{L}_{\odot}}
\def\AU{\textrm{AU}}
\def\hr{\textrm{hr}}
\def\dayUnit{\textrm{day}}

\def\Rsun{R_{\odot}}
\def\Dsun{D_{\odot}}
\def\thetasun{\theta_{\odot}}
\def\Omegasun{\Omega_{\odot}}
\def\alphasun{\alpha_{\odot}}
\def\decsun{\delta_{\odot}}
\def\deltasun{\delta_{\odot}}

\def\rVecsun{\hat{r}_{\odot}}

\def\Dearth{D_{\oplus}}
\def\betaearth{\beta_{\oplus}}
\def\xiearth{\xi_{\oplus}}
\def\alphaearth{\alpha_{\oplus}}
\def\deltaearth{\delta_{\oplus}}
\def\Dmoon{D_{\rm Moon}}

\def\Dsundiff{D_{\odot_{\rm diff}}}
\def\Dsunspec{D_{\odot_{\rm spec}}}

\def\tres{t_{\rm res}}
\def\tflash{t_{\star}}
\def\tspin{t_{\rm spin}}
\def\texp{t_{\rm exp}}
\def\tStar{t_{\star}}
\def\talign{t_{\rm align}}
\def\tmu{t_{\mu}}
\def\tvis{t_{\mu}^{\prime}}
\def\tfov{t_{\rm FoV}}

\def\muspin{\mu_{\rm spin}}
\def\Pphase{P_H}
\def\Porient{P_{\rm orient}}

\def\xsun{x_{\odot}}
\def\ysun{y_{\odot}}
\def\zsun{z_{\odot}}
\def\xearth{x_{\oplus}}
\def\yearth{y_{\oplus}}
\def\zearth{z_{\oplus}}

\def\wsun{w_{\odot}}
\def\ustar{u_{\star}}
\def\vstar{v_{\star}}
\def\wstar{w_{\star}}
\def\uearth{u_{\oplus}}
\def\vearth{v_{\oplus}}
\def\wearth{w_{\oplus}}

\def\deltastar{\delta_{\star}}
\def\alphastar{\alpha_{\star}}
\def\thetastar{\theta_{\star}}
\def\Hstar{H_{\star}}

\def\Fflash{F_{\star}}
\def\Rstar{R_{\star}}

\def\Vspec{{\cal V}_{\rm spec}}
\def\Vdiff{{\cal V}_{\rm diff}}

\def\Dbar{\overline{D}}
\def\Ddiff{D_{\oplus_{\rm diff}}}
\def\Dspec{D_{\oplus_{\rm spec}}}

\def\Aproj{A_{\rm proj}}
\def\thetares{\theta_{\rm res}}
\def\Deltav{\Delta V}
\def\thetafov{\theta_{\rm FoV}}
\def\OmegaL45{\Omega_{L4+L5}}
\def\Hedge{H_{\rm edge}}
\def\Omegafov{\Omega_{\rm FoV}}

\def\Ssun{{\cal S}_{\odot}}
\def\Searth{{\cal S}_{\oplus}}

\def\dprime{\prime\prime}

\def\NHigh{N_1}
\def\NLow{N_2}

\def\icarus{Icarus}
\def\PS1{{\rm PS1}}
\def\LSST{{\rm LSST}}

\def\ga{\gtrsim}
\def\la{\lesssim}

\def\endash{\text{--}}

\newcommand{\mean}[1]{\ensuremath{\langle #1 \rangle}}

\shorttitle{Glints \& SETI}
\shortauthors{Lacki}

\begin{document}

\title{A Shiny New Method for SETI: Specular Reflections from Interplanetary Artifacts}
\author{Brian C. Lacki$^{1}$}
\affiliation{Breakthrough Listen, Astronomy Department, University of California, Berkeley, CA, USA}
\email{astrobrianlacki@gmail.com}

\begin{abstract}
Glints of light from specular reflection of the Sun are a technosignature of artificial satellites.  If extraterrestrial intelligences have left artifacts in the Solar System, these may include flat mirror-like surfaces that also can glint.  I describe the characteristics of the resulting flashes.  An interplanetary mirror will appear illuminated for several hours, but if it is rotating, its glint may appear as a train of optical pulses.  The resulting glints can be very bright, but they will be seen only if the mirror happens to reflect sunlight to the Earth.  The detection of large mirrors is limited mainly by the fraction oriented to reflect sunlight toward Earth.  I give rough calculations for the expected reach of each exposure of Pan-STARRS1, LSST, and Evryscope for mirror glints.  A single exposure of Pan-STARRS1 has an effective reach of $10^{-9} \endash 10^{-7}\ \AU^3$ for interplanetary mirrors with effective areas of $10\ \meter^2$, depending on rotation rate.  Over several years, Pan-STARRS1 might accumulate a reach $\sim 10^5$ times greater than this, as it tiles the sky and different mirrors enter and exit a favorable geometry.
\end{abstract}


\section{Introduction}
Almost from the beginning of the Search for Extraterrestrial Intelligence (SETI), it's been suggested that aliens could travel across interstellar distances to our Solar System and leave behind artifacts \citep{Bracewell60,Sagan63}.  The occasional efforts to actively find these local material technosignatures, known as the Search for Extraterrestrial Artifacts (SETA), have received scant attention \citep{Freitas85}.  In recent times, SETA is growing in visibility, because of a reevaluation in the plausibility of starfaring and recent concepts of inorganic, space-dwelling intelligence and interstellar probes \citep{Dick03,Rose04,Tough04,Cirkovic06,Lubin16,Gertz16}.  Previously postulated SETA technosignatures include: (1) artifacts occupying special regions in the Solar System, like the Earth-Moon Lagrange points \citep{Freitas80,Freitas83,Steel95}; (2) radio emissions from small bodies that might actually be probes \citep{Enriquez18,Tingay18,Harp19}; (3) waste heat from power plants in the Solar System \citep{Papagiannis85}; (4) artificial light from artifacts or dwellings \citep{Loeb12}; (5) obvious material artifacts on the surfaces of Solar System bodies \citep{Arkhipov96-Moon,HaqqMisra12,Davies13}; and (6) identifiable fragments of technology mixed into the surface layers of Solar System bodies \citep{Arkhipov96-Meteor}.  No conclusive sign of extraterrestrial artifacts has been found \citep[e.g.,][]{deLaFuenteMarcos18}.

The purpose of this paper is to motivate another possible SETA signature: specular reflection of sunlight from the surfaces of artifacts in interplanetary space.  A flat mirror reflects sunlight into a narrow cone with the same angular extent as the Sun.  Since the Sun covers a very small part of the sky, specular reflections can be very bright, with an apparent flux falling as $\Dearth^{-2}$ instead of $\Dsun^{-2} \Dearth^{-2}$, where $\Dsun$ is the body's distance from the Sun and $\Dearth$ is its distance from Earth.  Bodies with liquid seas on their surfaces or ice crystals in their atmospheres can display specular reflection \citep{Sagan93,Williams08,Stephan10,Marshak17}, but oceans and atmospheres do not occur on small natural bodies.  Aside from a proposal in \citealt{Scheffer10} that we might observe specular reflection from solar panels on exoplanets, looking for flat mirrors does not appear to have been proposed in the SETA literature.
   
Satellite glints, of course, are a well-known phenomenon from human-made objects in Earth orbit, and thus are a proven technosignature.  During a satellite glint, sunlight is reflected off a mirror-like surface on a satellite, such as a solar panel.  Glints are a possible nuisance when searching for fast (milliseconds--seconds) optical transients \citep{Shamir06}.  They are one possible explanation for the Perseus Flasher transient in the 1980s \citep{Maley87,Schaefer87}, although some later studies disputed the prevalence of the glints \citep{Borovicka89,Varady92}.  In the 2000s, the Cherenkov telescope HESS performed a pilot study in looking for bright, fast transients.  It found one event, a triplet of flashes that can be explained as glints from a rotating satellite or fragment \citep{Deil09}.  Recently, \citet{Abeysekara16} reported the serendipitous discovery of a train of optical pulses crossing the VERITAS field of view during optical SETI observations of KIC 8462852, also attributed to a satellite.  Far more spectacular than any of these are the Iridium flares, with peak brightnesses exceeding magnitude $-7$, caused by the reflection of sunlight on the antennas of satellites in the old Iridium constellation \citep{Maley03}.\footnote{Most of the original Iridium constellation has been de-orbited, although at the present time $25$ failed Iridium satellites should remain in orbit, according to \citet{Sladen19}.}  

Any artifacts from extraterrestrial intelligences (ETIs) in low Earth orbit would surely be vastly outnumbered by human satellites at this point, but we have sent very few things beyond the orbit of the Moon.  If we did see a glint from elsewhere in the Solar System, particularly in a place we haven't sent anything, it would imply either a substantial population of shiny, ETI artifacts or a reflective surface deliberately aligned with the Earth, like a heliograph.  The glint would have the spectrum of sunlight, assuming a perfectly reflective surface.  Just like some of our own satellites, the artifact could be spinning.  The resultant glints would generally consist of a periodic train of pulses, as the reflection of the Sun repeatedly swept past the Earth.  Unlike glints from nearby objects, though, the spot size of these glints would be enormous, of order the size of the Sun for an object $1\ \AU$ or more away.  Thus the pulse train would last hours, not seconds, until its relative motion moved the Earth out of the beam.  Likewise, the glint would be visible from the entire Earth, and not just a region a few kilometers wide \citep[c.f.,][]{Maley03}.

The geometry of the specular reflection of sunlight from objects is reviewed in Section~\ref{sec:Geometry}.  The apparent brightness and photometric properties of glints are discussed in Section~\ref{sec:Photometry}.  In Section~\ref{sec:Density}, I calculate how many randomly-aligned mirrors would be needed in the Solar System in order to be detected in a single exposure of some photometric surveys.  Section~\ref{sec:Conclusion} provides a basic summary of the paper.

\section{The Geometry of Glints}
\label{sec:Geometry}
\subsection{Alignment between the Earth, Sun, and Artifact}
\begin{figure}
\centerline{\includegraphics[width=9cm]{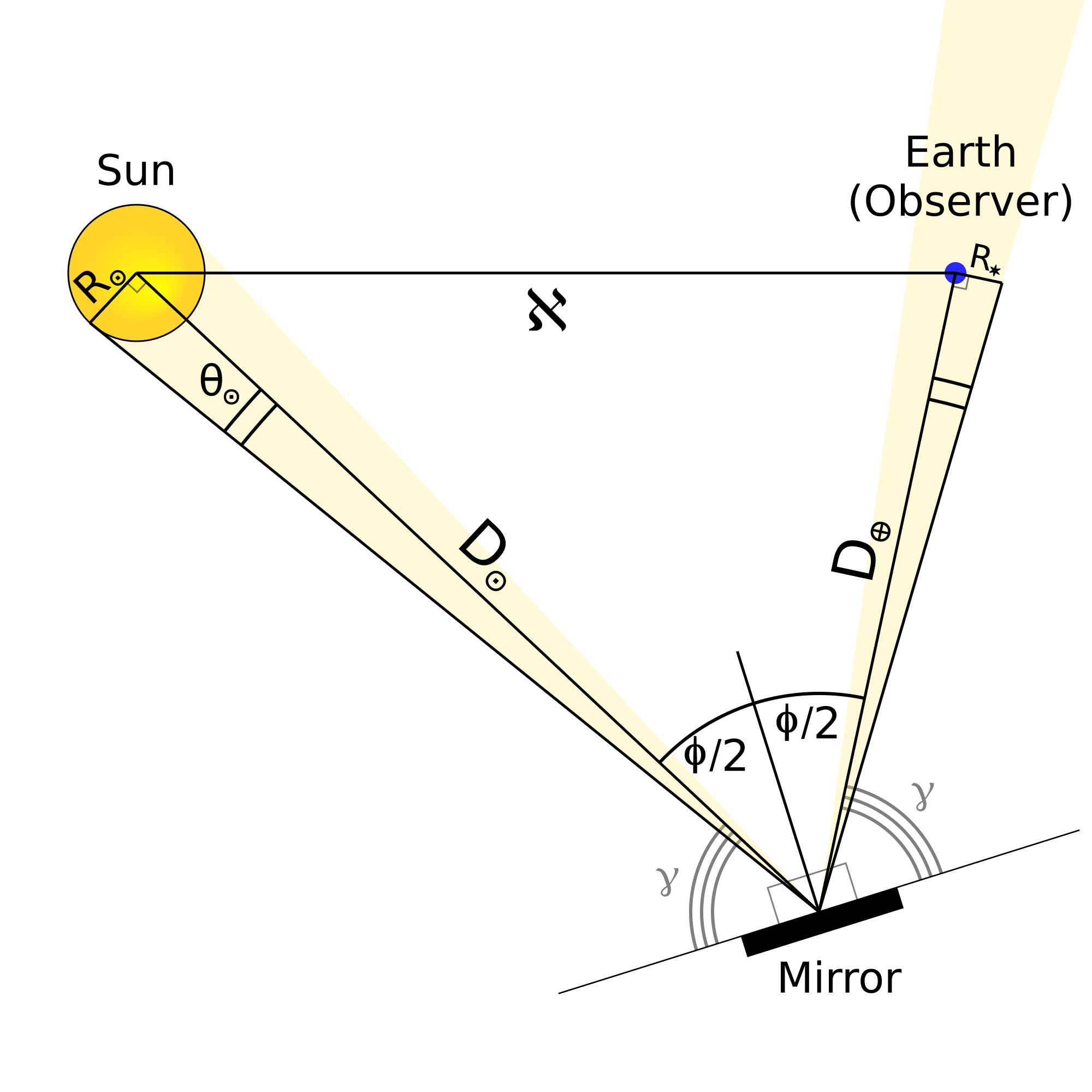}}
\figcaption{Sketch of the geometry between the mirror, the Sun and the Earth.\label{fig:SolarSystemGeometry}}
\end{figure}

The observable characteristics of specular reflections are fairly simple and arise directly from the geometry of the observer (assumed to be on the Earth for this paper), the illumination source (Sun), and the reflecting body.  A mirror is located at a distance $\Dsun$ from the Sun and $\Dearth$ from the observer (see Figure~\ref{fig:SolarSystemGeometry}).  In the vicinity of the mirror, the Sun appears as a disk with angular radius $\thetasun = \Rsun/\Dsun$, covering a solid angle of $\Omegasun = \pi (\Rsun/\Dsun)^2$.  The sunlight illuminating each infinitesimal area element of the surface is reflected into a cone with opening angle $\thetasun$ and covering solid angle $\Omegasun$.

If the illuminated surface as a whole is highly curved, then the cones of illuminated sunlight from each surface element point every which way.  The observable characteristics of an unresolved object is then similar to a diffusely reflecting surface.  Smaller curvatures concentrate the reflected sunlight into a smaller solid angle. Ideally, we'd view a completely flat surface (with curvature angle less than $\thetasun$), in which case all of the reflected sunlight is emitted into a solid angle $\Omegasun$.  For a randomly oriented two-sided flat mirror, the probability that it reflects sunlight towards the Earth is approximately $\thetasun^2 / [8 \cos(\phi/2)]$ when $\phi$, the mirror's phase angle, is not close to $\pi$, as demonstrated in Appendix~\ref{sec:Aim}.  As long as the sheet's dimensions are much smaller than $\Dearth$, but not so small that diffraction is important, then the reflected sunlight is visible within a spot radius $\Rstar = \thetasun \Dearth$ of the center-line of the cone.  The Rayleigh criterion gives a minimum size for a mirror that reflects sunlight into a solid angle $\sim \Omegasun$: $\ell_{\rm diff} = \lambda/\thetasun = 110\ \um\, (\lambda / 500\ \nm) (\Dsun / \AU)$.  Any mirror visible from interplanetary distances with current telescopes is much bigger.  

Both the observer and the artifact are moving and they have a relative velocity $\Deltav$.  The glint is visible as long as the observer, as viewed from the surface, is located within the cone of reflected sunlight.  On the surface's sky, the observer approaches within an angular impact parameter $\beta_{\oplus}$ of the center of the Sun's reflection.  Thus, the observer's path cuts an angular distance $\theta_{\star} \equiv 2 \sqrt{\thetasun^2 - \beta_{\oplus}^2} \equiv 2 f_{\beta} \thetasun$.  The glint remains visible for a time $\talign = \theta_{\star} \Dearth / \Deltav$:
\begin{equation}
\talign = \frac{2 f_{\beta} \Rsun \Dearth}{\Deltav \Dsun} = 12.9\ \hr\,f_{\beta} \left(\frac{\Deltav}{30\ \kms}\right)^{-1} \left(\frac{\Dearth}{\Dsun}\right).
\end{equation}
An artifact in the Moon's orbit ($\Dearth = 384,000\ \km$, $\Deltav = 1\ \kms$) would be visible for $1$ hour.  From the observer's perspective, the source is moving over the same angle $\theta_{\star}$, which is of order $0.5^{\circ}$ when $\Dsun \approx 1\ \AU$.   

Thus, these ``flares'' last for hours instead of the seconds-long flares from human satellites in low Earth orbit.  Because the sources are further away, the spot size of the glint is much larger -- $R_{\star} = \Rsun$ for $\Dearth = \Dsun$ -- and it would usually be visible from anywhere on Earth.

\subsection{Rotating mirrors}
\subsubsection{Qualitative effects of rotation}
While it's possible that there would be a steady flare lasting for $\talign$, glints may appear as a train of pulses.  That happens when the reflecting surface is rotating: the cone of reflected sunlight sweeps past the observer repeatedly during an encounter.  A derelict artifact is unlikely to have a stable pointing, and in general would spin with some period $\tspin$.

\begin{figure}
\centerline{\includegraphics[width=9cm]{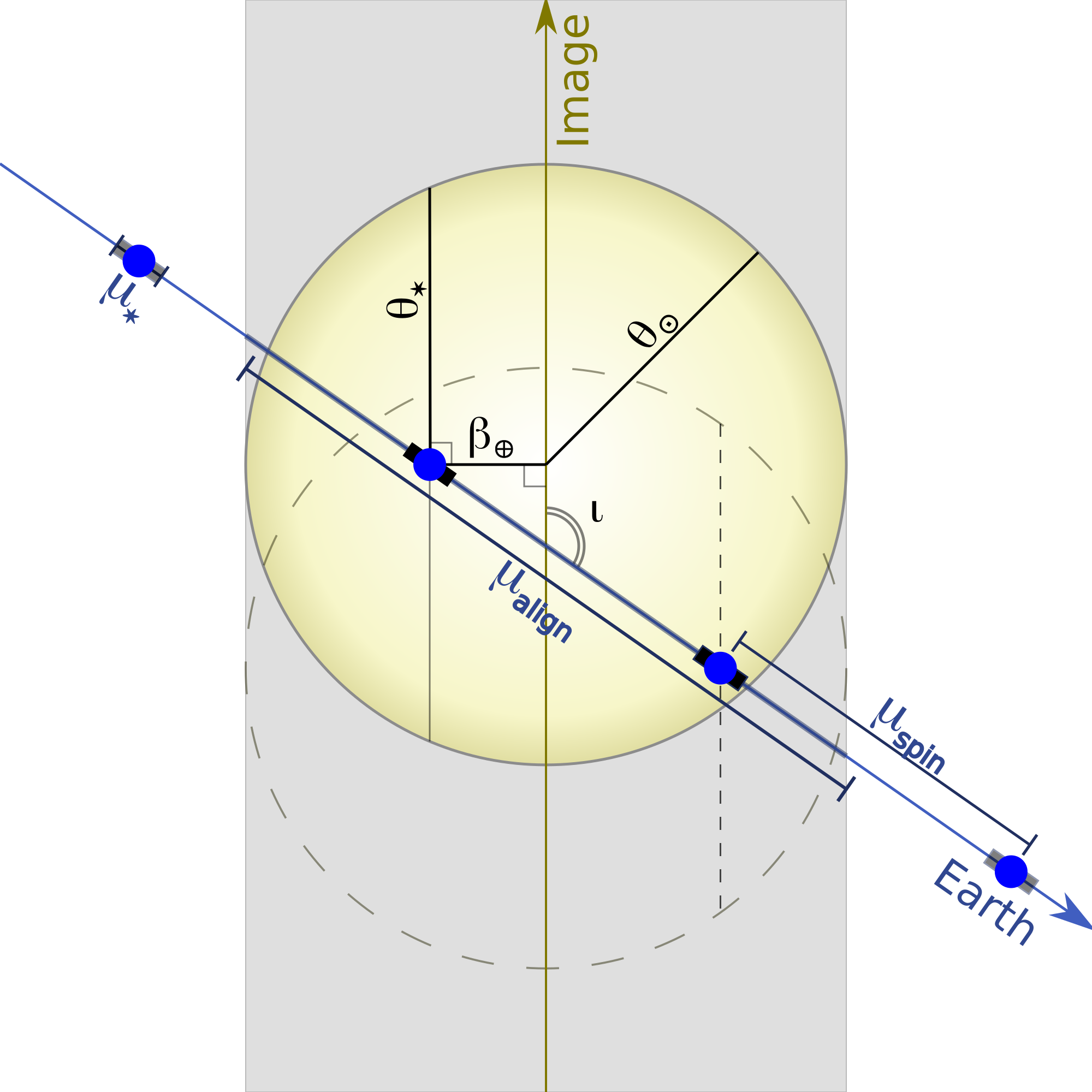}}
\figcaption{For a quickly rotating mirror, the image of the Sun usually sweeps across the sky in a constant path.  The Earth moves into this path and the image has a series of passages over the Earth each lasting $t_{\star}$.  The image is situated here for the midpoint of the closest passage, with the outline of its position during the midpoint of a second passage shown as a dashed line. \label{fig:PassageGeometry}}
\end{figure}

From the perspective of an observer on a rapidly rotating mirror surface, the Sun's image usually traces a path across the background stars during each rotation period.  The Earth slowly crawls into the path of the Sun where the image can ``cross'' it (Figure~\ref{fig:PassageGeometry}).  Now $\talign$ indicates the alignment between the Earth and the image's path on the sky.  If the surface is rapidly spinning ($\tspin \ll \talign$), there will be multiple passages, each sampling a different impact parameter $\betaearth$.  The smallest $\betaearth$ is drawn from a roughly uniform distribution from $0$ to $\muspin / (2 \sin \iota)$, where the angular distance traveled by the observer per rotation is $\muspin = \Deltav \tspin / \Dearth$ and $\iota$ is the angle between the local path of the revolving image and the Earth.  Thus, for a rapidly spinning mirror, the Earth passes very close to the center of the Sun's image during one of its passages.  This is unlike the case of a stationary surface, for which the Sun's image is simply a spot on the sky that the observer can graze.

A rapidly spinning artifact, therefore, will display a series of flashes that increase and then decrease in duration as $\sqrt{1 - (2t/\talign)^2}$, where $t$ is the time since closest passage.  The longest flash duration will be $\sim (\thetasun / \pi) \tspin \approx 0.0015 (\Dsun / \AU)^{-1} \tspin$.  In addition, only the Sun's limb is reflected during the earliest and latest flashes.  Limb darkening will affect their brightness and color.

\subsubsection{Path of the Sun's image}
\label{sec:SunPath}
\begin{figure*}
\centerline{\includegraphics[width=17cm]{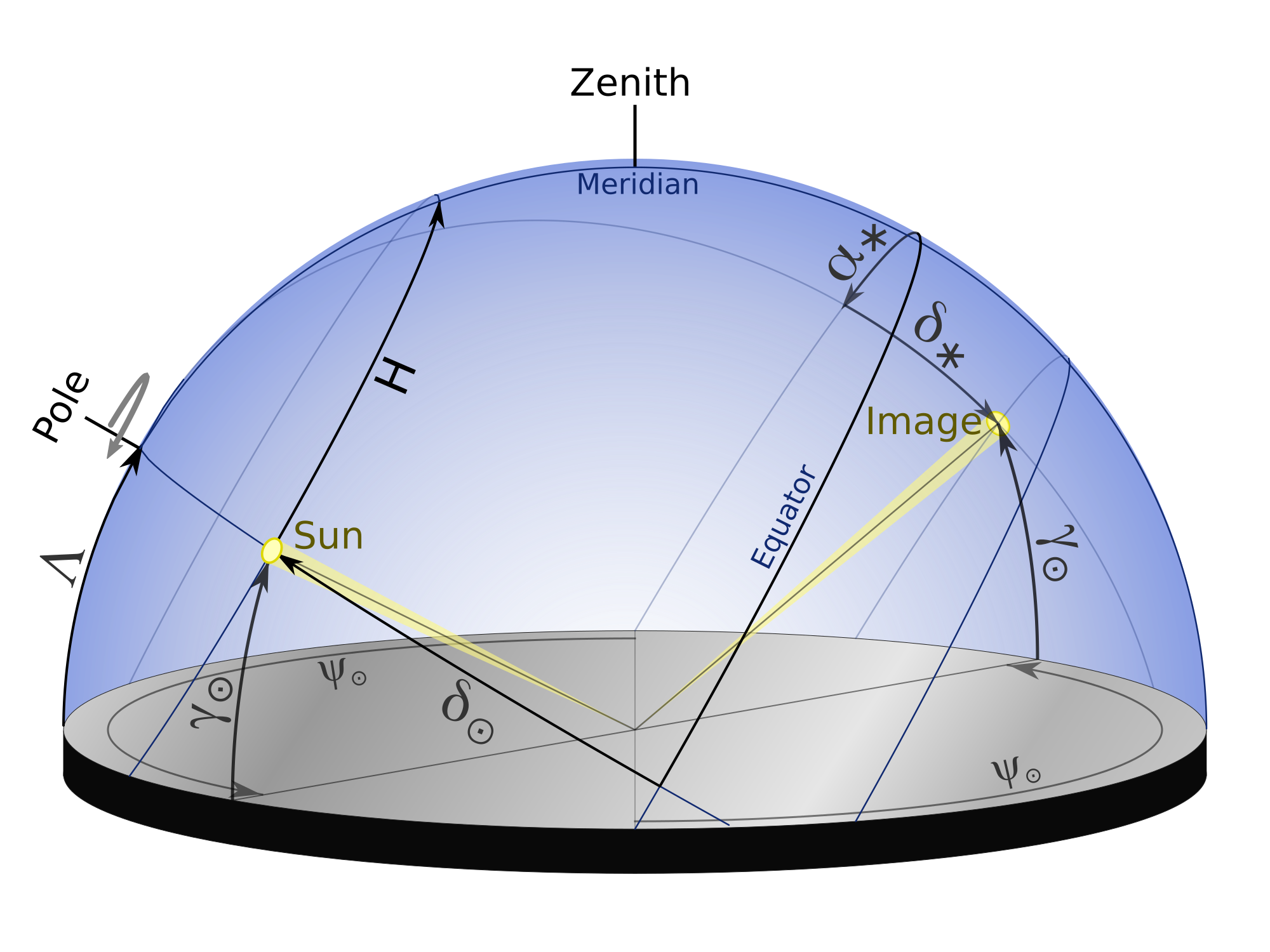}}
\figcaption{For a rotating mirror, coordinate systems analogous to right-ascension and declination, and altitude and azimuth, can be defined.  An observer on the surface of the mirror would find the sky has a celestial pole at altitude $\Lambda$.  At any given instant, the Sun is defined to have RA of zero, and declination $\deltasun$.  The image's RA $\alphastar$ and declination $\deltastar$ change also with the hour angle $H$.\label{fig:MirrorSkyGeometry}}
\end{figure*}

Geometrically, the artifact's rotation can be likened to the Earth's rotation.  The flat reflecting surface can be imagined to be simply one parcel of the surface of a much larger rotating sphere.  From the mirror's perspective, the sky revolves around the spin axis, with two fixed ``poles'' (Figure~\ref{fig:MirrorSkyGeometry}).  Thus, a coordinate system analogous to right ascension ($\alpha$) and declination ($\delta$) can be used for the rotating sheet.  It is also clear that the altitude of the celestial ``pole'' for the surface defines an angle isomorphic to latitude ($\Lambda$).  As on the Earth, the right ascension along the meridian progressively increases during each ``day'' caused by rotation, allowing us to parameterize rotation with an hour angle ($H$).  

The Sun's location on the mirror's sky, ($\alphasun$, $\deltasun$), can be converted into an altitude $\gamma_{\odot}$ and azimuth $\psi_{\odot}$.  The reflection of the Sun is aimed at the opposite azimuth ($\psi_{\star} = \pi + \psi_{\odot}$) and same altitude ($\gamma_{\star} = \gamma_{\odot}$) as the Sun.  We see the reflection, centered at ($\alphastar$, $\deltastar$), if it overlaps with the Earth's position on the sky ($\alpha_{\oplus}$, $\delta_{\oplus}$).  Since right ascension's zero point is arbitrary, I choose $\alphasun = 0$ throughout this paper.  Appendix~\ref{sec:ImagePath} derives the path of the Sun's image on the mirror sky.  These paths can include ``retrograde'' motion.  

Broadly speaking, there are two regimes for the reflection's behavior.  In some rare cases, the spin axis is nearly coincident with the normal to the mirror surface, with $|\Lambda| = \pi/2$.  This is the polar regime, where the image of the Sun is fixed against the sky.  If the image also coincides with the Earth, the glint is visible but does not flash despite the rotation.  The chance of this happening for a randomly oriented mirror is approximately $\thetasun^4 / 32$.  It's vastly more likely for a mirror to be observed in a subpolar regime, with $|\Lambda| \le \pi/2 - \thetasun/2$.  In these cases, the mirror rotation causes the Sun's image to travel against the background sky, resulting in flashes from our perspective.

\subsubsection{Accurate Flash Duration for Mirrors at Opposition}

Suppose the Earth is located at phase angle $\phi$ from the Sun in the artifact's sky.  A glint will be visible from Earth if the reflection of the Sun comes within $\thetasun$ of the Earth's position on the sky at some point during a rotation period.  As long as $\talign \gg \tspin$, the duty cycle of each flash is usefully parameterized by a range of hour angles during one spin period.  It depends on the ``latitude'' of the glint surface, the ``declination'' of the Sun, and the position of the Earth on the artifact's sky:
\begin{equation}
\Hstar = \int_{H \in [0, 2\pi), |(\alpha_{\rm spec}, \delta_{\rm spec}) - (\alpha_{\oplus}, \delta_{\oplus})| < \thetasun} dH.
%\Hstar = \int_{\begin{array}{l} H \in [0, 2\pi) \\ |(\alpha_{\rm spec}, \delta_{\rm spec}) - (\alpha_{\oplus}, \delta_{\oplus})| < \thetasun \end{array}} dH.
\end{equation}
For most orientations of the artifact, the glint never intersects with the Earth and $\Hstar = 0$.  Otherwise, each flash of the glint lasts for $\tflash = \Hstar \tspin / (2 \pi)$.  In general, this is $\sim \thetasun \tspin$ but if the ``latitude'' is nearly $\pm \pi/2$, $\tflash \approx \tspin$ is possible.

The values of $\cos H$ where the glint begins and ends can be found using the quartic equation derived in Appendix~\ref{sec:ImageEarthAngle}.  A simpler expression for $\Hstar$ exists if the artifact happens to be at opposition with respect to the Earth ($\phi = 0$).  Then the Earth coincides with the Sun, and $\Hstar$ measures the fraction of each ``day'' where sunlight is reflected back at the Sun's center.\footnote{Where the Earth (observer) is essentially a point against the much larger Sun.}  That occurs, for a two-sided mirror, when the Sun is at the zenith or the nadir, and a prerequisite for that happening is $|\deltasun \pm \Lambda| \la \thetasun / 2$ (when $\thetasun \ll 1$).  $\Hstar$ is the same for $\deltasun = \Lambda + \Delta$ and $\deltasun = -\Lambda - \Delta$:
\begin{equation}
\label{eqn:HStarOpposition}
\Hstar = 2 \cos^{-1} \min\left[\frac{\cos (\thetasun/2) - \sin \Lambda \sin \deltasun}{\cos \Lambda \cos \deltasun}, -1\right].
\end{equation}

\section{Photometry}
\label{sec:Photometry}
\subsection{The flux from a glint}
When an observer is in the cone of reflected sunlight, the entire reflecting surface will appear to light up with the Sun's surface brightness, reduced by a specular reflectivity $\rho$.  During a flash, the instantaneous flux from the mirror is
\begin{equation}
\label{eqn:FlashFlux}
\Fflash = \frac{\rho \Lsun \Aproj}{4 \pi^2 R_{\sun}^2 \Dearth^2}.
\end{equation}
The quantity $\Aproj = A \cos (\phi/2)$ is the projected area of the mirror from the Earth, and $\Lsun$ is the luminosity of the Sun.  This equation applies to the bolometric flash brightness for a wavelength-independent $\rho$, and for narrowband fluxes with $\rho$, $\Fflash$, and $\Lsun$ replaced by their values for a particular bandpass.  For comparison, the flux from a surface reflecting diffusely and uniformly into the sky (Lambertian reflection) with albedo $\rho$ is:
\begin{equation}
F_{\rm diff} = \frac{\rho \Lsun \Aproj}{4 \pi^2 \Dsun^2 \Dearth^2} = \left(\frac{\Rsun}{\Dsun}\right)^2 \Fflash = 2.2 \times 10^{-5} \left(\frac{\Dsun}{1\ \AU}\right)^{-2} \Fflash.
\end{equation}

Artifacts would probably be spinning, visible for only a small fraction of the time as the cone of reflected sunlight repeatedly sweeps the Earth.  The instantaneous flux may be a useful quantity for some setups with very fast photodetectors, like the Cherenkov telescopes that occasionally see satellite glints \citep{Deil09,Abeysekara16}.  CCD based surveys, though, which make up the majority of large-sky surveys, have exposure times from seconds to minutes.  As noted above, a typical flash will last only $\sim 0.1 \%$ of a rotation.  A glint may be visible for only a short fraction of one exposure, and disappear for the next several exposures.

Another factor a survey has to contend with is the motion of the artifact on the sky.  Surveys like Pan-STARRS have a high enough angular resolution $\thetares$ that the glint will appear to move on the sky between exposures.  Thus, exposures cannot be simply co-added to deepen the sensitivity to these glints -- rather than remaining a stationary point, the artifact will leave a dotted trail on the sky.  The object may be smeared even in individual exposures, if it is closer than:
\begin{equation}
\label{eqn:Dblur}
D_{\rm blur} = 2.48\ \AU \left(\frac{\Deltav}{30\ \kms}\right) \left(\frac{\texp}{1\ \min}\right) \left(\frac{\thetares}{1\ \arcsec}\right)^{-1
} .
\end{equation}
There are methods that could recover a streak that is too faint to be seen in individual spots by summing the power along the entire trail, as commonly used in near Earth asteroid searches \citep[e.g.,][]{Waszczak17}.  I will be conservative, though, and assume that a glint can only be seen if it can be detected before moving out of one resolution element for a survey.  Furthermore, I will assume that the detection algorithm can only be run on only one image, and that there is no co-addition of images even if the artifact moves slowly enough not to smear.

\begin{deluxetable*}{lccc}
\tablewidth{0pt}
\tablecolumns{4}
\tablecaption{Duty cycle dependence on timescales\label{table:GlintDutyCycle}}
\tablehead{\colhead{Order} & \colhead{$\eta$} & \colhead{$N_{\rm flash}$} & \colhead{Description}}
\startdata
$\tmu \le \tflash, \texp$                    & $\tmu / \texp$                  & $1$                & High proper motion limits integration\\
$\texp \le \tflash, \tmu$                    & $1$                             & $1$                & Very short exposure resolves flash\\
$\tflash \le \tmu \le \tspin, \texp$         & $\tflash / \texp$               & $1$                & High proper motion, only one unresolved flash\\
$\tflash \le \texp \le \tmu, \tspin$         & $\tflash / \texp$               & $1$                & Short exposure, only one unresolved flash\\
$\tflash \le \tspin \le \tmu \le \texp$      & $\tflash \tmu / (\tspin \texp)$ & $\tmu / \tspin$    & Many flashes integrated, limited by motion\\
$\tflash \le \tspin \le \texp \le \tmu$      & $\tflash / \tspin$              & $\texp / \tspin$   & Many flashes integrated, limited by exposure
\enddata  
\end{deluxetable*}

Now the average flux during an exposure is given by
\begin{equation}
\mean{\Fflash} = \Fflash \eta.
\end{equation}
There are four time scales relevant for calculating the duty cycle fraction $\eta$: (1) the duration $\tflash$ of an individual flash, (2) the rotation period $\tspin$ of the artifact, (3) $\texp$, the exposure time, (4) the time $\tres = \thetares \Dearth / \Deltav$ it takes for the artifact to cross one resolution element (of angular width $\theta_{\rm res}$) in the image, and (5) $\talign$, the time the train of flashes lasts.  Since the latter two matter only in that they limit how long a mirror can be observed, they can be combined into $\tmu = \min(\tres, \talign)$, a timescale describing the effects of proper motion.  The duty cycle depends on the relative sizes of these timescales.  Table~\ref{table:GlintDutyCycle} gives approximations for $\eta$ for exposures where at least one flash is observed.  Roughly, $\eta$ is the fraction of an exposure a single flash lasts multiplied by the number of flashes $N_{\rm flash}$ seen during the exposure.  Diffusely reflecting objects also have a duty cycle, $\eta_{\rm diff} \approx \min(\tres/\texp, 1)$, accounting for the effects of proper motion on sensitivity:
\begin{equation}
\label{eqn:etaDiff}
\mean{F_{\rm diff}} = F_{\rm diff} \eta_{\rm diff}.
\end{equation}

\subsection{Detectability of a glint}
How far can we see these glints with our current instruments?  The answer depends on $\tspin$.  It also depends on telescope exposure time, angular resolution, field of view (FoV), and sensitivity.  Lower angular resolution actually increases $\tres$, since it takes longer for the artifact to cross a resolution element, but it also blends more sky background and makes it harder to see faint flashes.  Likewise, with a shorter integration time, each exposure is less sensitive, but those with a flash will have less dead time.

\begin{deluxetable*}{l|ccccc|cccccc|c}
\tablewidth{0pt}
\tablecolumns{13}
\tablecaption{Adopted parameters for considered surveys\label{table:Survey}}
\tablehead{\colhead{Property} & \multicolumn{5}{c|}{Pan-STARRS1 $3\pi$} & \multicolumn{6}{c|}{LSST} & \colhead{Evryscope}}
\startdata
Angular resolution          & \multicolumn{5}{c|}{$1.1\ \arcsec$}                             & \multicolumn{6}{c|}{$0.2\ \arcsec$}                                    & $26.6\ \arcsec$\\
Field radius                & \multicolumn{5}{c|}{$1.5^{\circ}$}                              & \multicolumn{6}{c|}{$1.75^{\circ}$}                                    & $52.5^{\circ}$\\
Instantaneous field of view & \multicolumn{5}{c|}{$7.1\ \deg^2$}                              & \multicolumn{6}{c|}{$9.6\ \deg^2$}                                    & $8660\ \deg^2$\\
Instrument reference        & \multicolumn{5}{c|}{\citet{Chambers16}}                         & \multicolumn{6}{c|}{\citet{LSST09}}                                    & \citet{Law15}\\
Photometry reference        & \multicolumn{5}{c|}{\citet{Tonry12}}                            & \multicolumn{6}{c|}{\citet{LSST09}}                                    & \citet{Bessell90}\\
\hline
Bands                       & $g_{\PS1}$ & $r_{\PS1}$ & $i_{\PS1}$ & $z_{\PS1}$ & $y_{\PS1}$ & $u_{\LSST}$ & $g_{\LSST}$ & $r_{\LSST}$ & $i_{\LSST}$ & $z_{\LSST}$ & $y_{\LSST}$ & V\\
Magnitude limit             &	$22.0$     & $21.8$     & $21.5$     & $20.9$     & $19.7$     & $23.9$      & $25.0$      & $24.7$      & $24.0$        & $23.3$      & $22.1$      & $16.4$ \\
Solar absolute magnitude    & $-26.6$    & $-26.9$    & $-27.1$    & $-27.0$    & $-27.0$    & $-25.4$     & $-26.5$     & $-26.9$     & $-27.0$       & $-27.1$     & $-27.0$     & $-26.7$\\
Exposure time               & $43\ \sec$ & $40\ \sec$ & $45\ \sec$ & $30\ \sec$ & $30\ \sec$ & $15\ \sec$  & $15\ \sec$  & $15\ \sec$  & $15\ \sec$   & $15\ \sec$  & $15\ \sec$  & $120\ \sec$ 
\enddata
\tablecomments{Solar absolute magnitude is the apparent magnitude of the Sun as viewed from 1 AU in each filter, using the transmission curves for each band and instrument.}
\end{deluxetable*}

I consider three nominal surveys (Table~\ref{table:Survey}).  The first, Pan-STARRS1, represents an extant, deep survey, with deep exposures lasting for tens of seconds, and arcsecond angular resolution \citep{Chambers16}.  These capabilities will all be sharpened further with the Large Synoptic Survey Telescope (LSST) when it comes online in a few years, which is the second survey I consider \citep{LSST09}.  Both of them have instantaneous fields of view of less than ten square degrees, although they image over a thousand square degrees during the course of a night.  Artifacts far in the Solar System are aligned with the Earth for hours, so there is a fair chance of capturing it.  Neither presents an opportunity for capturing more than a couple of exposures before the glint fades permanently, though, so it would be hard to verify the glint's properties.  Since Pan-STARRS1 and LSST image in multiple bands, for each set of parameters and survey, I select the band that yields the strongest limits when presenting results.

Evryscope, the third program considered, serves as a notional instantaneously all-sky survey.  This proposed system is an array of telescopes, covering $9,000$ square degrees instantaneously \citep{Law15}.  It will image this entire field every two minutes, so any glint bright enough to be seen can be tracked and verified to be visible while it moves across an angle $\thetasun$ on the sky.  Evryscope has moderate sensitivity, down to 16th magnitude, and sub-arcminute angular resolution.

\begin{figure*}
\centerline{\includegraphics[width=9cm]{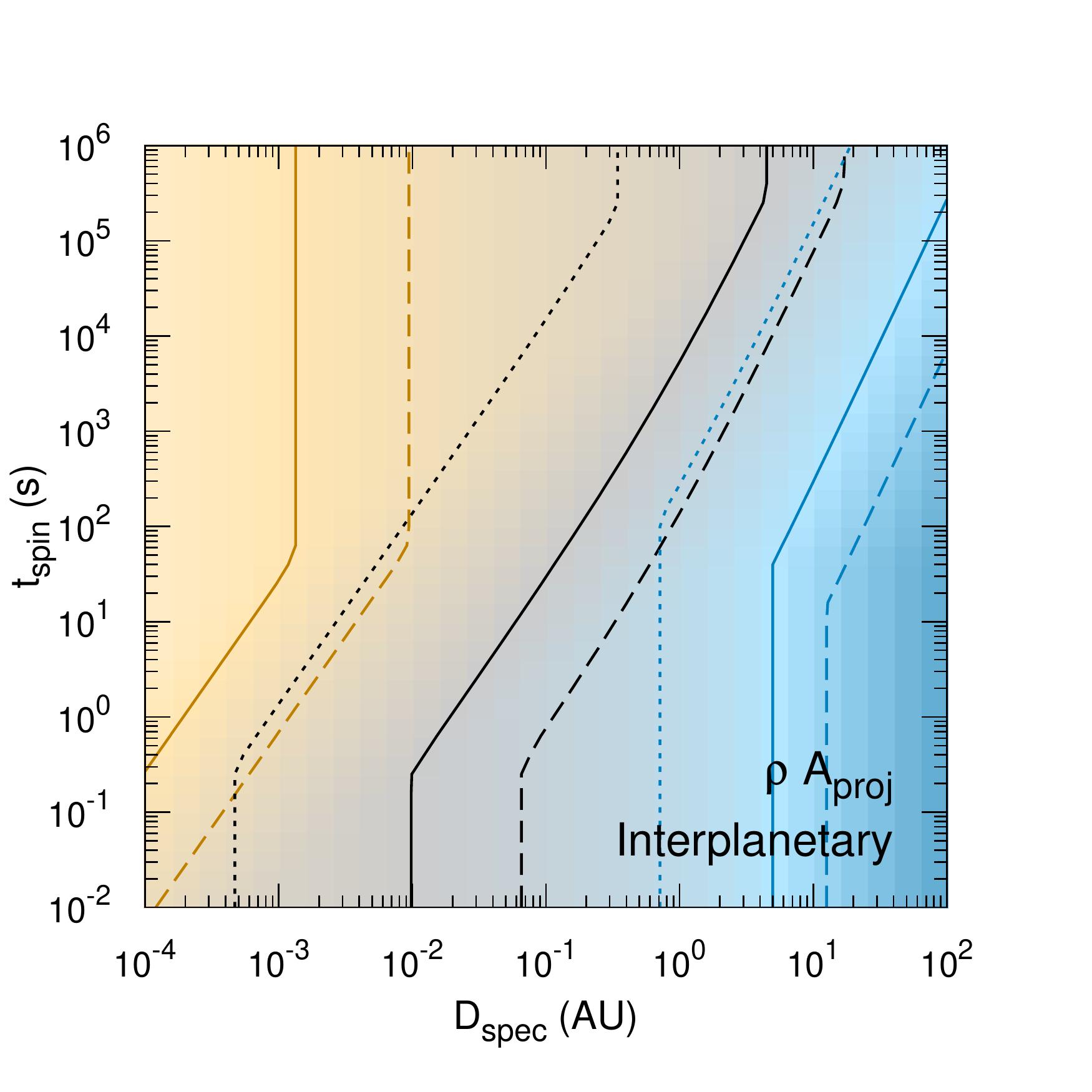}\includegraphics[width=9cm]{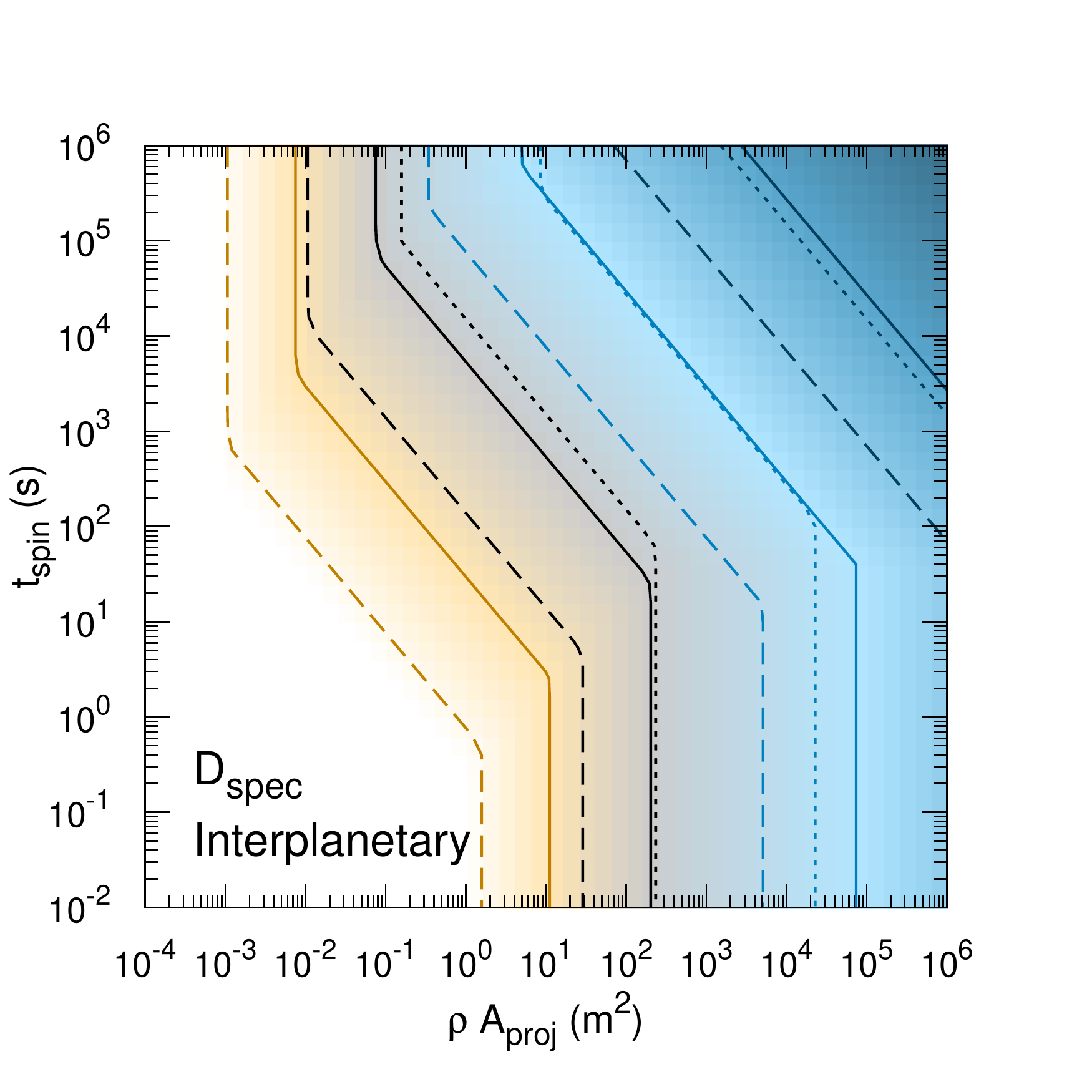}}
\figcaption{Left: Area of a specularly reflecting surface at opposition and moving at $\Delta v = 30\ \kms$ that can be seen by various photometric surveys, as a function of its rotation frequency $\nu_{\rm spin}$ and distance. The effective area is $(100\ \meter)^2$ for cyan lines, $1\ \meter^2$ for black lines, and $1\ \cm^2$ for orange lines.  Right: Maximum distance a surface can be seen ($\Delta v = 30\ \kms$) as a function of $\nu_{\rm spin}$ and area.  Contours are for ranges of $0.1$ (orange), $1$ (black), $10$ (light cyan), and $100\ \AU$ (dark cyan).  Surveys are Pan-STARRS (solid lines and shading), LSST (dashed), and Evryscope (dotted).  In both plots, I assume $\Hstar = \thetasun$.  \label{fig:GlintRange}}
\end{figure*}

Figure~\ref{fig:GlintRange} plots the range that mirrors are visible, assuming $\Deltav = 30\ \kms$ as appropriate for an interplanetary artifact..  We see that the sensitivity usually decreases as the artifact spins faster, even for $\tspin$ as long as $10^5\ \sec \sim 1\ \dayUnit$ when the range is 1 AU.  The range usually falls by a factor of $\sim 1,400$ between the asymptotic limits for slow and fast rotation.  This transition regime in rotation period changes with mirror area and the survey used.  With Pan-STARRS1 and $\rho \Aproj = 1\ \meter^2$, the slow rotation regime occurs for $\tspin \ga 400,000\ \sec$, while fast rotation regime occurs for $\tspin \la 0.3\ \sec$.  Large objects are in the fast rotation regime when $\tspin \la 40\ \sec$ ($\sim \texp$) and do not achieve the slow rotation regime even when $\tspin = 10^6\ \sec$.  Small, closer surfaces need to be spinning much more rapidly for rotation to affect the photometry because they cross the sky quicker and sensitivity is limited by proper motion instead. 

If it weren't for rotation, Pan-STARRS1 (LSST) could detect a $1\ \meter^2$ surface out to about the distance of Jupiter (Uranus).  A rotation period of a minute drastically reduces the range.  For $\tspin = 100\ \sec$, Pan-STARRS1 (LSST) has a range of just $0.2\ \AU$ ($0.9\ \AU$), with a minimal range of $0.01\ \AU$ ($0.07\ \AU$) in the fast rotation regime.  Of course, larger reflecting surfaces are visible to greater distances.  Even by just increasing the area to $10\ \meter^2$, Pan-STARRS1 (LSST) could detect a quickly rotating mirror out to $0.1\ \AU$ ($0.5\ \AU$).  These results are exciting because the laser sails proposed for Breakthrough Starshot have an area of $10\ \meter^2$ \citep[e.g.,][]{Parkin18} and the solar sail demonstrator IKAROS has an area of $200\ \meter^2$ \citep{Tsuda11}.  Although neither is capable of entering orbit around another star, this demonstrates that human-scale artifacts are detectable with forthcoming surveys.  LSST should be able to detect a mirror of that size in Venus' or Mars' solar orbit at closest approach, if the mirror orientation is favorable. 

\section{Possible constraints on artifact density}
\label{sec:Density}

\subsection{Reach of photometric surveys}
Two quantities determine whether a glint is visible from Earth: the probability that it is observed during an exposure and its mean brightness during the exposure.

\begin{deluxetable*}{lcc}
\tablewidth{0pt}
\tablecolumns{3}
\tablecaption{Phase probability dependence on timescales\label{table:Pphase}}
\tablehead{\colhead{Order} & \colhead{$\Pphase$} & \colhead{Description}}
\startdata
$\tvis \le \tflash, \tspin, \texp$          & $\tflash / \tspin$                 & High proper motion samples just one phase\\
$\texp  \le \tflash, \tspin, \tvis$         & $\tflash / \tspin$                 & Short exposure samples just one phase\\
$\tflash \le \texp \le \tspin, \tvis$       & $\texp / \tspin$                   & Phase range limited by exposure\\
$\tflash \le \tvis \le \tspin, \texp$       & $\tvis / \tspin$                   & Phase range observed limited by proper motion\\
$\tflash \le \tspin \le \tvis, \texp$       & $1$                                & All phases of mirror observed during exposure
\enddata  
\end{deluxetable*}

Even if a mirror is in the field of view, and even if the Earth is in the revolving path of the Sun's image on the rotating mirror's sky, that doesn't mean a flash will be seen during an exposure.  We also need to consider $\Pphase$, the probability that we observe the mirror at the right phase (the right time of its day) to see a flash at some point during the integration time.  Now this probability depends on $\tflash$, $\tspin$, and $\texp$, but it also depends on the duration the mirror is in the field of view and reflects glints at the observer, $\tvis = \min(\talign, \tfov)$.  Here, $\tfov \approx \thetafov \Dearth / \Delta v$, where $\thetafov$ is the radius of the survey's instantaneous field of view.  Approximations for $\Pphase$ are listen in Table~\ref{table:Pphase}.  

We can now calculate the effective volume sampled by a single exposure:
\begin{multline}
\label{eqn:Veff}
\Vspec = \int_{\rm FoV} \int_{-1}^1 \int_{-1}^1 \int_{D_{\rm min}}^{D_{\rm max}} \max \left(\frac{\texp}{\tvis}, 1\right) \Pphase \Dearth^2 \\
\times \frac{d^2 \Porient}{d\cos \deltasun d\cos \Lambda} d\Dearth d\cos \deltasun~ d\cos \Lambda~ d\Omega.
\end{multline}
$D_{\rm max}$ is the maximum distance a glint is visible from a mirror with a fixed size, rotation period, $\deltasun$, $\Lambda$, and direction relative to Earth.  Each part of the exposure sky area $\Omega$ corresponds to different values of $\phi$ and $\xiearth$ during a flash, where the latter is the ``position angle'' of the Earth relative to the Sun on the artifact.  Thus, $\Hstar$ varies across the integration range of $\Omega$.  This integral also includes a dependence on the orientation of the artifact and its spin axis.  For randomly oriented surfaces,
\begin{equation}
\frac{d^2 \Porient}{{d\cos \deltasun}{d\cos \Lambda}} = \frac{1}{4}.
\end{equation}
The enhancement factor $\max \left(\texp/\tvis, 1\right)$ matters for very long exposures, as new artifacts move into the field of view or start to glint while the old ones are no longer visible.\footnote{Of the two basic geometry regimes, a subpolar one where $\Hstar \ll 2 \pi$ and a polar one $\Hstar \approx 2 \pi$, the subpolar one is vastly more important for $\Vspec$.  Even though a polar geometry lets the glint stay aligned with Earth for the entire rotation period, letting it be viewed from further distances, the probability of a polar geometry is proportional to $\thetasun^4$.  A subpolar glint only requires $|\deltasun \pm \Lambda| \la \thetasun / 2$.}

These equations are very complicated in the most general case of arbitrary phase angle $\phi$.  The integrals are easier to calculate if the artifact happens to be at opposition with respect to the Earth ($\phi = 0$).  Then $\Hstar$ is given by equation~\ref{eqn:HStarOpposition} and $\Aproj = A$.  

When the exposure sky areas is small relative to $\thetasun^2$, a mirror near opposition has $\phi \approx 0$ and an irrelevant $\xi_{\oplus}$.  Thus, there is then no dependence on $\Omega$ in Equation~\ref{eqn:Veff}.  This assumption is violated for the cameras of many of the large surveys that might hope to detect glints, including PAN-STARRS1, LSST, and Evryscope.  To get a rough sense of the effective volume of these surveys, I nonetheless use this opposition approximation for the entire field of view.  

\begin{figure*}
\centerline{\includegraphics[width=9cm]{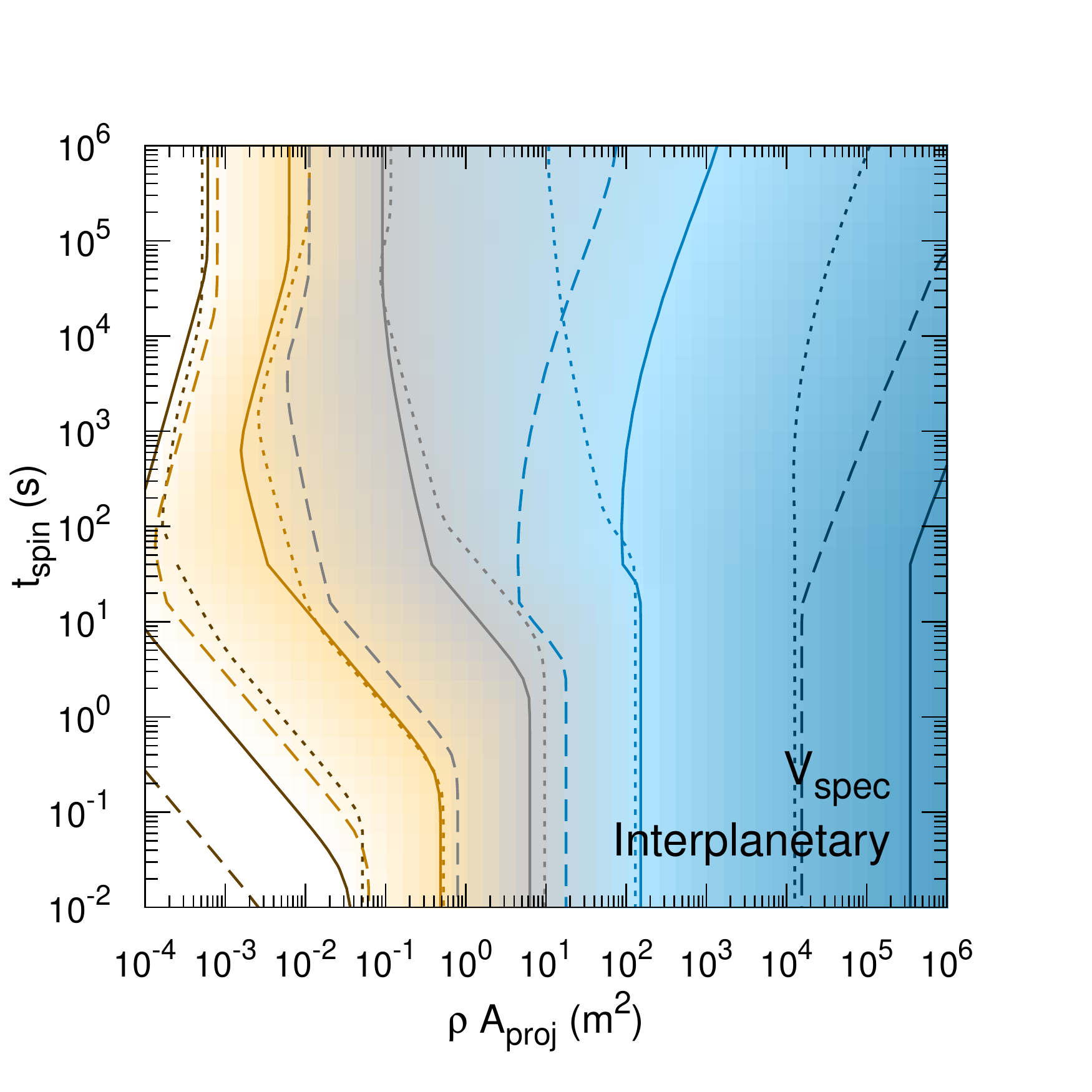}\includegraphics[width=9cm]{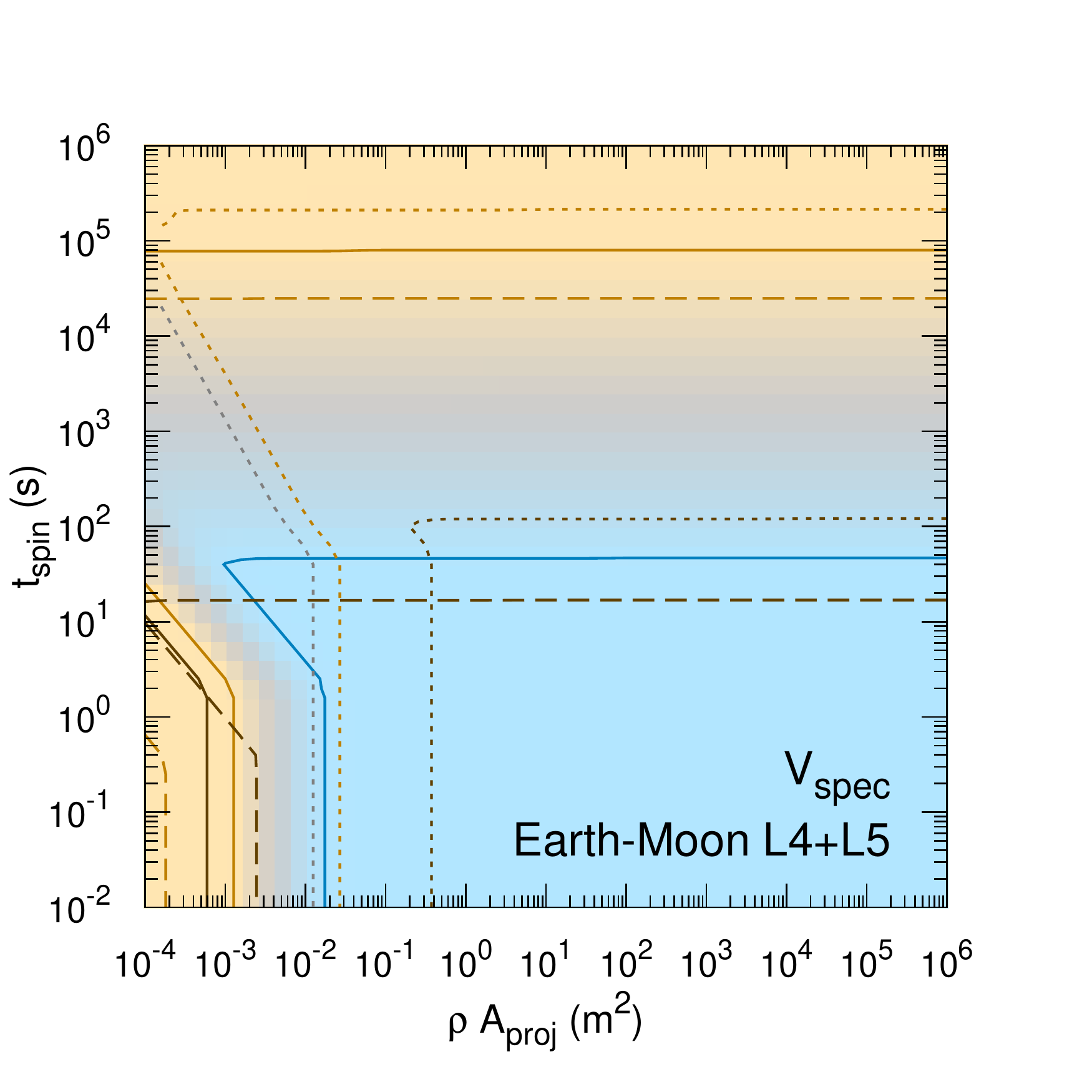}}
\figcaption{Effective survey volumes for artifacts in interplanetary space (left, $\Deltav = 30\ \kms$) and near the Earth-Moon L4/L5 points (right, $\Deltav = 1\ \kms$) using the opposition approximation.  On left, contours are for $10^{-3}$ (dark cyan), $10^{-6}$ (light cyan), $10^{-9}$ (grey), $10^{-12}$ (light orange), and $10^{-15}\ \AU^3$ (dark orange).  On right, the contours denote limits on the number of mirrors in the Earth-Moon L4/L5 points: $10 \NHigh$ (dark orange), $\NHigh$ (light orange), and $\NLow$ (cyan).  Surveys use the same line styling as Figure~\ref{fig:GlintRange}. \label{fig:VeffConstraints}}
\end{figure*}

Figure~\ref{fig:VeffConstraints} is a plot of the resulting $\Vspec$ for PAN-STARRS1 and Evryscope.  I consider interplanetary artifacts with $\Deltav = 30\ \kms$ ($D_{\rm min} = 10^{-4}\ \AU$, $D_{\rm max} \le 1,000\ \AU$).  I also consider mirrors trapped in the Earth-Moon L4 and L5 points, with $\Deltav = 1\ \kms$.  The integration region for the Lagrange points is defined using the dimensions of the stable halo orbits quoted in \citet{Freitas83}: $|\Dearth - \Dmoon| \le 75,000\ \km$ with a maximum sky area of $\OmegaL45 = 2 \times 300,000 \times 70,000\ \km^2 / \Dmoon^2$ substituting for the field of view of Evryscope, where I adopt $\Dmoon = 384,000\ \km$ as the distance to the Moon.

When dealing with interplanetary artifacts (left panel), the effective survey volume is small -- $0.001\ \AU^3$ for square kilometer mirrors, $10^{-9} \endash 10^{-7}\ \AU^3$ for ten square meter mirrors -- both because of the narrow fields of view of the deeper Pan-STARRS1 and LSST, and the stringent constraints on mirror alignment for a glint to be seen.  For large mirrors, the aim of the mirror and the field of view is critical since even a distant glint can be seen easily.  This is why, for large mirrors, Evryscope is more powerful than even LSST.  It's also why there's a greater reach when the big mirrors rotate quickly, since the glint sweeps across a larger part of the Solar System during an exposure.  For small mirrors, in contrast, better sensitivity comes from better depth per exposure.  In this case, a small $D_{\rm max}$ is likely to lead to streaking of the glint on the exposure (equation~\ref{eqn:Dblur}), further weakening sensitivity.

If we look for mirrors near the Earth-Moon L4/L5 points (right panel), they are so slow and so high flux that whether the mirror aims sunlight at Earth during an exposure is basically the determining factor.  In mirrors with very slow rotation, the probability one will reflect sunlight to the Earth in the opposition approximation is $\thetasun^2/8$, leading to this upper limit on the number of mirrors in the Lagrange point regions:
\begin{equation}
\NHigh = \frac{\OmegaL45}{\Omegafov} \frac{8}{\thetasun^2} = \frac{8 \OmegaL45}{\Omegafov} \left(\frac{\Dsun}{\Rsun}\right)^2 .
\end{equation} 
Quickly rotating mirrors are more likely to emit a flash towards the Earth, with probability $\varepsilon \thetasun / (4 \pi)$, where I find $\varepsilon \approx \pi^2$ in my calculations.  Then the upper limit on the number of mirrors in the Lagrange regions is:
\begin{equation}
\NLow = \frac{\OmegaL45}{\Omegafov} \frac{4 \pi}{\varepsilon \thetasun} \approx \frac{4 \OmegaL45}{\pi \Omegafov} \frac{\Dsun}{\Rsun}.  
\end{equation}
In fact, these two limits define the behavior over most of the considered parameter space in Figure~\ref{fig:VeffConstraints}, with $\NLow$ holding for $\tspin \la \texp$, and $\NHigh$ holding for $\tspin \ga 2,000\ \texp$, regardless of surface area.  Only when looking for the smallest mirrors (smaller than $0.05\ \meter^2$ for Pan-STARRS1) does survey depth matter, and the constraints become weaker than the pure geometric expectation.

Of course, during a night, Pan-STARRS1 and LSST observe thousands of square degrees.  Since $\talign$ is about the length of one night, if an artifact is glinting somewhere within a given night's observing region, there's a good chance the image beam will still be passing over the Earth by the time these telescopes catch up to that region.  For order of magnitude estimates, we can regard the effective instantaneous field of view for these instruments to be the distinct sky area covered in one night.  Thus, Pan-STARRS and LSST have a nightly effective reach that is $\sim 100$ times greater than the reach per exposure.  This raises Pan-STARRS1's effective reach for $10\ \meter^2$ mirrors to $\sim 10^{-7} \endash 10^{-5}\ \AU^3$ per night, for instance. 

Over longer periods, the shiny artifacts are cycled through observability by the constantly changing alignment between them, the Sun, and the Earth.  After each period of $\talign$, glints from a new population of objects become visible while the old glints become misaligned.  Since $\talign$ is roughly half a day, each night will provide a new sample of phase space.  Furthermore, it is unlikely that an artifact will glint on two different nights over periods much shorter than the synodic period.  This will allow long-running programs like LSST to further tighten constraints by factors of hundreds.  Accurately modeling the rate that new mirrors become visible is difficult, however, because the phase angle depends not only on the ecliptic longitudes of the mirror and Earth, but the orientation of the mirror itself.  An actual calculation is deferred, but as an order of magnitude estimate, this sampling effect can increase the yield by at most $2\pi/\thetasun \approx 1,400 (\Dsun/1\ \AU)$.   In all, Pan-STARRS1 and LSST might achieve an effective reach about $\sim 10^5$ times greater than that of a single exposure.  The final reach of Pan-STARRS1 to $10\ \meter^2$ mirrors is roughly $10^{-4} \endash 10^{-2}\ \AU^3$.

\subsection{Comparing artifact detectability by diffuse and specular reflections}
The density constraints on shiny artifacts, as shown in Figure~\ref{fig:VeffConstraints}, are actually quite weak.  In fact, for most values of $\rho \Aproj$ and $\tspin$, large diffusely reflecting objects are better constrained by all three surveys, when considering single exposures.  The effective volume of a survey for diffusely reflecting surfaces is 
\begin{multline}
\label{eqn:Vdiff}
\Vdiff = \int_{\rm FoV} \int_{-1}^1 \int_0^{D_{\rm max}} \max \left(\frac{\texp}{\tfov}, 1\right) \Dearth^2 \frac{dP}{d\cos \phi} \\
\times d\Dearth d\cos \phi d\Omega,
\end{multline}
where $dP/d\cos\phi = (1/2)$, and the projected area is calculated as $\Aproj = A \cos (\phi/2)$.  I used equation~\ref{eqn:Vdiff} to calculate these effective volumes for each of the three surveys, after substituting $\mean{F_{\rm diff}}$ (equation~\ref{eqn:etaDiff}) for the flux to account for proper motion.  

\begin{figure}
\centerline{\includegraphics[width=9cm]{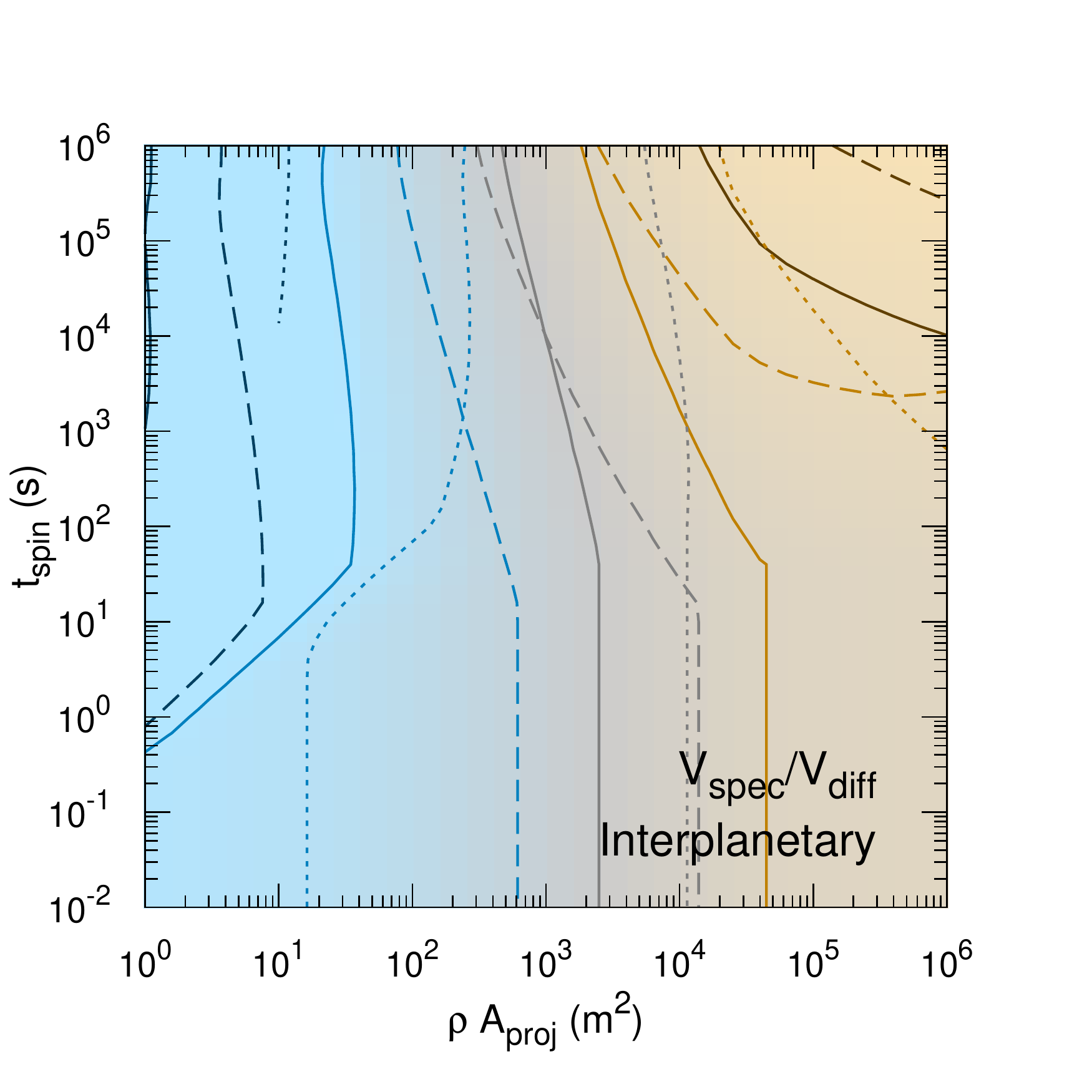}}
\figcaption{Ratio of the effective survey volume for mirrors and diffusely reflecting surfaces with $\Deltav = 30\ \kms$.  The contours stand for $\Vspec / \Vdiff$ of $10^6$ (dark cyan), $10^3$ (light cyan), $1$ (grey), $10$ (light orange), and $100$ (dark orange).  I use the opposition approximation.  Surveys use the same line styling as Figure~\ref{fig:GlintRange}.\label{fig:nRatios}}
\end{figure}

Figure~\ref{fig:nRatios}~shows how the ratios of $\Vspec / \Vdiff$ varies with surface area and rotation period for the mirrors.  It is easier to detect a large surface when it is diffusely reflecting rather than a pure specular reflection.  For small objects, specular reflections are much more detectable than the diffuse reflections, with $\Vspec / \Vdiff = 1$ for effective areas of $\sim 500 \endash 2,500\ \meter^2$ for Pan-STARRS1 ($\sim 100 \endash 600\ \meter^2$ for LSST and $10,000\ \meter^2$ for Evryscope).  Slow rotation slightly decreases this area.  Two factors are at work: (1) only a small fraction of shiny objects are visible because most are pointed the wrong way and (2) when objects are further from the Sun, they must be more precisely aligned for us to see a glint.

Consider the non-rotating case, in the dominant subpolar regime, with negligible proper motion.  Flat mirrors are detectable within
\begin{equation}
\Dspec \approx \sqrt{\frac{\Lsun}{4\pi^2 \Rsun^2} \frac{\rho \Aproj}{F_{\rm min}}},
\end{equation}  
while a dull artifact is visible out to
\begin{equation}
\Ddiff \approx \sqrt{\frac{\Lsun}{4\pi^2 \Dsundiff^2} \frac{\rho \Aproj}{F_{\rm min}}},
\end{equation}
where $\Dsundiff$ is the distance between the Sun and the dull artifact.  The shiny artifact is visible at much larger distances than the dull artifact.  Non-rotating shiny artifacts have a probability
\begin{equation}
P = \frac{1}{4} \left(\frac{\Rsun}{\Dsunspec}\right)^2
\end{equation}
of aiming their glint towards the observer, where $\Dsunspec$ is the distance between the Sun and the shiny artifact.  The ratio of instantaneous effective volume surveyed for specular and diffuse reflections is:
\begin{equation}
\label{eqn:NonRotatingVComparison}
\frac{\Vspec}{\Vdiff} \approx P \left(\frac{\Dspec}{\Ddiff}\right)^2 = \frac{1}{4} \frac{\Dsundiff}{\Rsun} \left(\frac{\Dsundiff}{\Dsunspec}\right)^2.
\end{equation}

If we are looking for very small artifacts, they will be visible only when they are close to Earth.  Thus, each $\Dearth$ is much smaller than $\Dsun \approx 1\ \AU$, and equation~\ref{eqn:NonRotatingVComparison} comes out to
\begin{equation}
\frac{\Vspec}{\Vdiff} \approx \frac{1}{4} \frac{\Dsun}{\Rsun} \approx 50,
\end{equation}
greatly favoring shiny artifacts.  If $A$ has a medium value, $\Dspec \ga 1\ \AU$ but $\Ddiff \ll 1\ \AU$.  The shiny artifact is generally further from the Sun, with $\Dsunspec \approx \Dspec$, requiring more precise aim towards the Earth:
\begin{equation}
\frac{\Vspec}{\Vdiff} \approx \frac{1}{4} \frac{\aleph^3}{\Rsun \Dspec^2} \approx 50 \left(\frac{\Dsunspec}{\AU}\right)^{-2}.
\end{equation}
Finally, a large artifact is visible all the way out to the outer Solar System, with $\Ddiff \approx \Dsundiff \ga 1\ \AU$:
\begin{equation}
\Ddiff \approx \left(\frac{\Lsun}{4\pi^2}\frac{A}{F_{\rm min}}\right)^{1/4} \equiv \Dbar,
\end{equation}
and the advantage of shiny over dull artifacts is flipped:
\begin{equation}
\frac{\Vspec}{\Vdiff} \approx \frac{1}{4} \frac{R_{\sun}}{\Ddiff} \approx \frac{1}{900} \left(\frac{\Ddiff}{\AU}\right)^{-1}.
\end{equation}

\begin{deluxetable*}{lccccc}
\tablewidth{0pt}
\tablecolumns{6}
\tablecaption{Approximations for ratio of effective reach for shiny and dull surfaces in selected regimes\label{table:ReachScalings}}
\tablehead{\colhead{Regime} & \colhead{Rotation} & \colhead{Proper motion} & \colhead{$\Ddiff \ll \Dspec \ll \aleph$} & \colhead{$\Ddiff \ll \aleph \ll \Dspec$} & \colhead{$\aleph \ll \Ddiff \ll \Dspec$}}
\startdata
$\texp \le \tres, \tfov, \tflash, \tspin, \talign$          & Slow     & Slow     & $\displaystyle \frac{\aleph}{\Rsun}$ & $\displaystyle \frac{\Rsun \aleph^3}{\Dbar^4}$ & $\displaystyle \frac{\Rsun}{\Dbar}$\\
$\tres \le \texp \le \tfov, \tflash, \tspin, \talign$       & Slow     & Moderate & $\displaystyle \left(\frac{\aleph}{\Rsun}\right)^4$ & $\displaystyle \left(\frac{\Deltav \texp}{\thetares}\right)^2 \frac{\aleph^6}{\Dbar^8}$ & $1$\\
$\tres \le \tfov \le \texp \le \tflash, \tspin, \talign$    & Slow     & Fast     & $\displaystyle \left(\frac{\aleph}{\Rsun}\right)^2$ & $\displaystyle \left(\frac{\Deltav \texp}{\thetares}\right)^2 \frac{\Rsun^2 \aleph^4}{\Dbar^8}$ & $\displaystyle \left(\frac{\Deltav \texp}{\thetares} \frac{\Rsun^3}{\Dbar^4}\right)^{2/3}$ \\
$\tflash \le \texp \le \tspin, \tres, \tfov, \talign$       & Moderate & Slow     & $\displaystyle \left(\frac{\tspin}{\texp} \frac{\aleph}{\Rsun}\right)^{1/2}$ & $\displaystyle \left(\frac{\texp}{\tspin} \frac{\Rsun \aleph^9}{\Dbar^{10}}\right)^{1/3}$ & $\displaystyle \left(\frac{\texp}{\tspin} \frac{\Rsun}{\Dbar}\right)^{1/3}$\\
$\tflash \le \tres \le \texp \le \tspin, \tfov, \talign$    & Moderate & Moderate & $\displaystyle \left(\frac{\tspin \texp^5}{\thetares^6} \frac{\Deltav^6 \aleph^7}{\Rsun \Dbar^{12}}\right)^{1/2}$ & $\displaystyle \left(\frac{\Deltav^9 \texp^{10}}{\thetares^9 \tspin} \frac{\Rsun \aleph^{18}}{\Dbar^{28}}\right)^{1/3}$ & $\displaystyle \left(\frac{\Deltav^3 \texp^4}{\thetares^3 \tspin} \frac{\Rsun}{\Dbar^4}\right)^{1/3}$\\
$\tres \le \tflash \le \texp \le \tspin, \tfov, \talign$    & Moderate & Moderate & $\displaystyle \frac{\texp}{\tspin} \left(\frac{\aleph}{\Rsun}\right)^5$ & $\displaystyle \frac{\Deltav \texp^2}{\thetares \tspin} \frac{\aleph^6}{\Rsun^3 \Dbar^4}$ & $\displaystyle \frac{\thetares}{\Deltav \tspin} \frac{\Dbar^4}{\Rsun^3}$\\
$\tflash \le \tspin \le \texp \le \tres, \tfov, \talign$    & Fast     & Slow     & $\displaystyle \left(\frac{\aleph}{\Rsun}\right)^{1/2}$ & $\displaystyle \left(\frac{\Rsun \aleph^9}{\Dbar^{10}}\right)^{1/3}$ & $\displaystyle \left(\frac{\Rsun}{\Dbar}\right)^{1/3}$\\
$\tflash \le \tspin \le \tres \le \texp \le \tfov, \talign$ & Fast  & Moderate & $\displaystyle \left(\frac{\aleph}{\Rsun}\right)^2$ & $\displaystyle \left(\frac{\Deltav \texp}{\thetares}\right)^2 \frac{\aleph^6}{\Dbar^8}$ & $1$\\
$\tflash \le \tres \le \tspin \le \texp \le \tfov, \talign$ & Fast  & Moderate & $\displaystyle \left(\frac{\Deltav^6 \tspin^3 \texp^3}{\thetares^6} \frac{\aleph^7}{\Rsun \Dbar^{12}}\right)^{1/2}$ & $\displaystyle \left(\frac{\Deltav^9 \tspin^2 \texp^7}{\thetares^9} \frac{\Rsun \aleph^{18}}{\Dbar^{28}}\right)^{1/3}$ & $\displaystyle \left(\frac{\Deltav^3 \tspin^2 \texp}{\thetares^3} \frac{\Rsun}{\Dbar^4}\right)^{1/3}$
\enddata
\tablecomments{Scalings assume that both specularly and diffusely reflecting objects are in the same regime.  Rotation is slow if $\texp \le \tflash$, moderate if $\tflash \le \texp \le \tspin$, and fast if $\tspin \le \texp$.  Proper motion is slow if $\texp \le \tres$, moderate if $\tres \le \texp \le \tfov$, and fast if $\tfov \le \texp$.  $\Dbar = (\Lsun A_{\rm eff} / (4 \pi^2 F_{\rm min}))^{1/4}$.  $\aleph$ is the distance between the observer and the Sun, assumed to be $1\ \AU$.  Constant coefficients are ignored.}
\end{deluxetable*}

Rotation and proper motion affect the scalings in different ways (Table~\ref{table:ReachScalings} for selected cases).  Specular reflections always are more likely to be detected than diffusely reflecting objects of the same size and rotation period ($\Vspec / \Vdiff > 1$) when the surfaces are small and nearby ($\Dearth \ll 1\ \AU$).  Rotation alone tends to moderate the ratio of $\Vspec/\Vdiff$, pushing it closer to $1$ for all combinations of $\Dearth$.  This is because a rotating object has a smaller mean brightness but sweeps the image of the Sun across a larger sky area.  Proper motion is more severe for closer objects, and diffusely reflecting surfaces must be closer to be detected than a mirror.  Thus, rapid proper motion greatly favors specular reflections, and when it is fast enough $\Vspec / \Vdiff$ can reach $1$ even for distant objects.  Since proper motion is a limiting factor for deep photometric surveys, it is common for $\Vspec / \Vdiff \gg 100$ for small objects (Figure~\ref{fig:nRatios}).  Extremely rapid proper motion, with $\tfov \ll \texp$, moderates the advantage of specular reflections somewhat, since diffusely reflecting objects have a higher enhancement factor as more cross the field of view during an exposure.

\section{Conclusions}
\label{sec:Conclusion}
The glint of sunlight from a shiny artifact is a novel technosignature for SETI, demonstrated by its occurrence from human satellites in Earth orbit.  Glints are visible from a larger distance than a dull object, but only a small fraction of artifacts is visible at any one time.  Specular reflections from lightweight mirrors might act like a heliograph, serving as a cheap ``beacon'' for ETIs who can deliver a payload to the Solar System.  These would be able to use the power provided in sunlight passively and would not need any kind of electronics.   The relative motion of the Earth, Sun, and artifact brings the glint in and out of alignment over a timescale of several hours.  We would see it as a small Solar System object that appears and vanishes on that same timescale.  

Rotating objects have more complicated light curves, but generally consist of a regular train of flashes with a typical duty cycle of a fraction of percent.  The train of flashes will likewise last for several hours.  There is a polar regime in which the Sun and the surface's axis of rotation are aligned, for which the glint appears steady, but they contribute negligibly to the visible population of glints.  Rotating objects have a smaller average brightness, but the likelihood of seeing one is higher.

I used an opposition geometry approximation to simplify analysis and calculated possible constraints on mirror density set by widefield photometric surveys.  Pan-STARRS1 is sensitive to surfaces at $1\ \AU$ with $A \approx 200\ \meter^2$ (rapidly spinning) to $0.07\ \meter^2$ (slowly spinning).  Likewise, it is sensitive to a $10\ \meter^2$ mirror out to a distance of $0.1\ \AU$ (rapidly spinning) to $\ga 10\ \AU$ (slowly spinning).  Despite these relatively large ranges, searches for specular reflections are severely hampered by how narrow the cone of reflected sunlight is.  The effective search volume probed by a single exposure of Pan-STARRS1 for that $10\ \meter^2$ mirror in interplanetary space is $3 \times 10^{-9}\ \AU^3$ (rapidly spinning) to $6 \times 10^{-8}\ \AU^3$ (slowly spinning), peaking at $10^{-7}\ \AU$ at $\tspin \approx 1,000\ \sec$. Generally, rapid rotation in a mirror increases our sensitivity for large mirrors but decreases our sensitivity for small mirrors.  The geometry factor makes it harder to constrain shiny mirrors than dull surfaces of the same effective area when they are large.  However, specular reflection would greatly enhance a small artifact's visibility, especially when it is close enough to be blurred by proper motion.

We would detect a glint in each exposure only if there are millions of mirrors in the inner Solar System.  The Earth-Moon Lagrange points might be more promising, where only a few hundred mirrors are necessary for a detection using a very widefield survey like Evryscope.  Of course, these surveys image thousands of square degrees per night, so their field of view is effectively hundreds of times larger.  In addition, they repeatedly image the sky over the course of years, allowing us to examine hundreds of different geometries.  A model of how the yield grows with time would be useful in further characterizing this technosignature.

\appendix
\section{How Precisely Must the Mirror Be Aimed?}
\label{sec:Aim}
The probability a mirror is instantaneously aimed so as to produce a glint seen from Earth is the major limitation to this technosignature in a lot of cases.

The beam of reflected sunlight fills a solid angle of $\Omegasun$.  Since this is true no matter which direction the mirror is pointed, we expect by conservation of phase space that when we average over all mirror orientations and all image positions on the sky, that the probability is $\Omegasun/(4\pi)$.  This is in fact correct when averaged over all orientations: if an isotropic spherical shell of tiny mirrors (reflective on both sides) with random orientations surrounded a free-floating observer at a distance $\ll \Dsun$, then a fraction $\Omegasun/(4\pi)$ of them would show a glint to the observer when averaged over all sky directions.

But this general argument does not apply to specific cases.  In the opposition approximation, where both the Sun and the Earth are coaligned, a glint is only seen if the Sun and Earth are normal to the surface.  Tilting the mirror by just $\thetasun/2$ ensures the Sun's image no longer touches the Earth.  If both sides of the mirror are reflective, the solid angle of allowed glint alignments is $\pi \thetasun^2 / 2$, just half the mean value.  

\begin{figure*}
\centerline{\includegraphics[width=9cm]{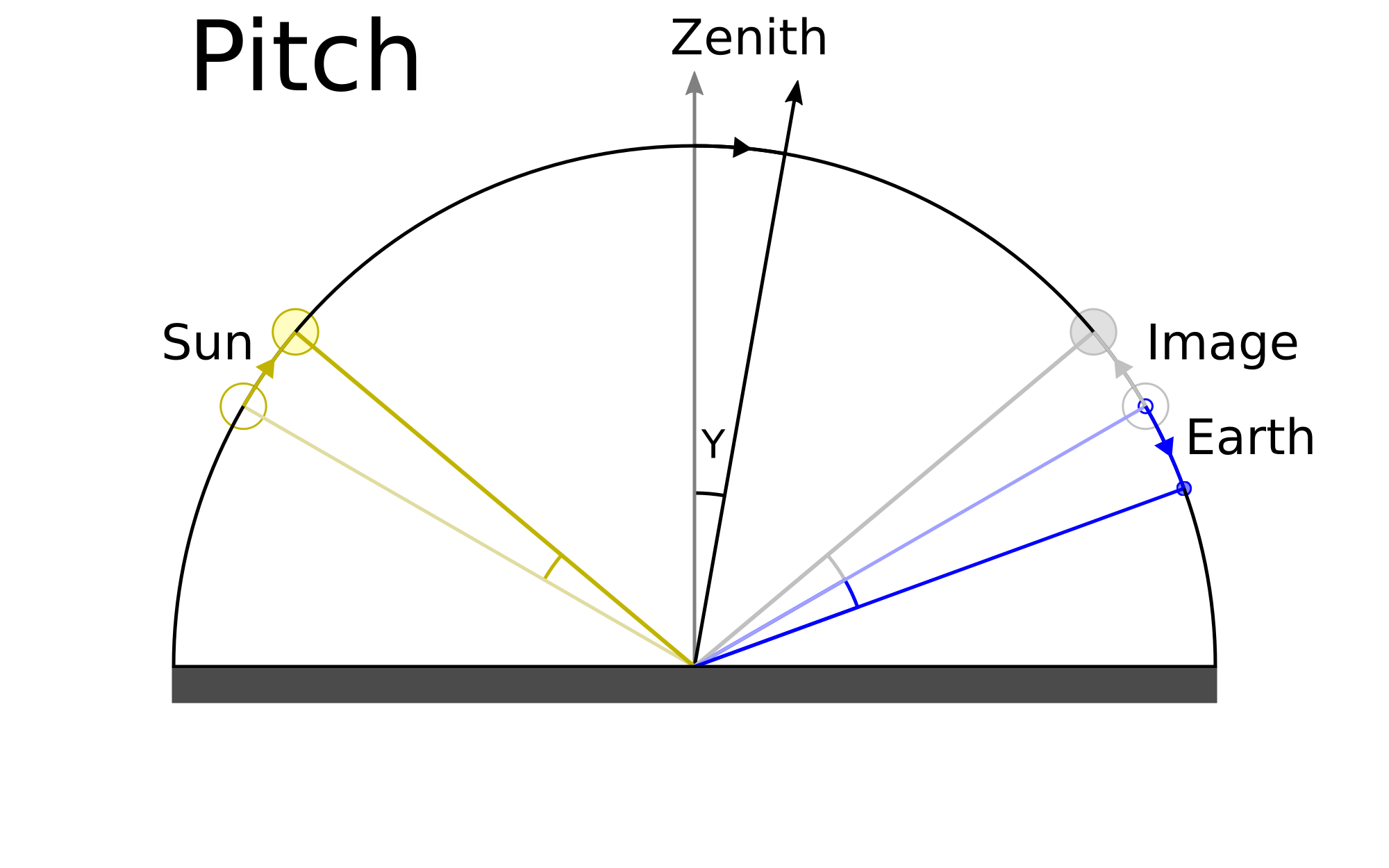}\includegraphics[width=9cm]{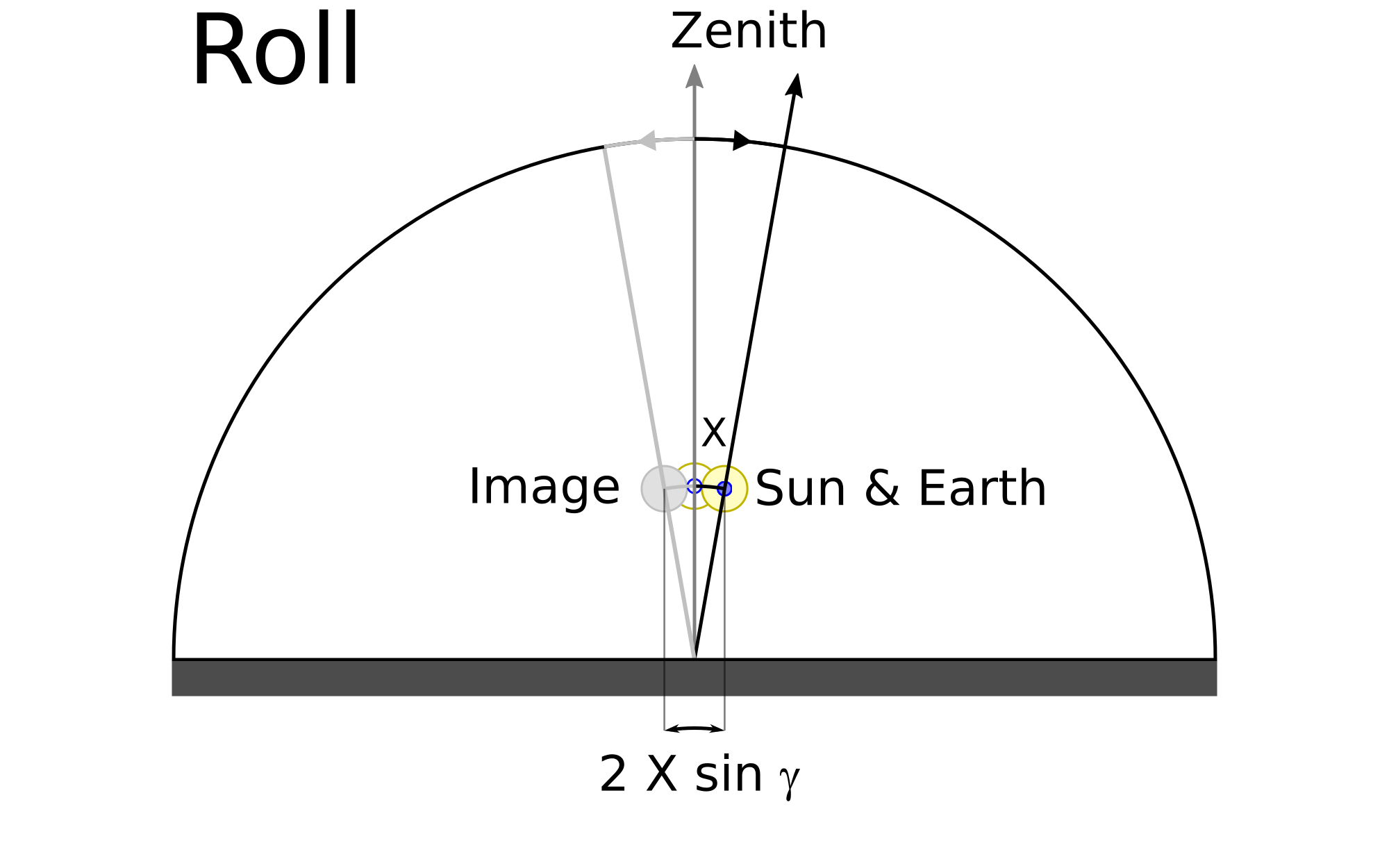}}
\figcaption{The angle a mirror can be tilted before a glint can no longer be seen depends on whether the tilting is a pitch (left) or a roll (right) motion.  Pitching the mirror by angle $Y$ leads to a widening by $2Y$ of the angular separation between the Earth and the image.  When rolling the mirror by angle $X$, the distance the Sun and Earth move on the sky is proportional to $\sin\gamma$, where $\gamma$ is the altitude.  In the right diagram, the Sun is on the far side of the celestial sphere, while the image and the Earth are on the near side.\label{fig:RotationTolerance}}
\end{figure*}

The key to understanding this discrepancy is understanding the different effects of rolling and pitching the mirror.  Let $\hat{z}$ be the direction normal to the mirror's surface.  The $\hat{x}$ direction will be defined to point in the direction of the Sun's azimuth, and the $\hat{y}$ direction is perpendicular to $\hat{x}$ and $\hat{z}$.  Now suppose the Sun's reflected image and the Earth are perfectly aligned.  Tilting the mirror will cause the Sun and the Earth will appear to move one way on the mirror's sky, while the Sun's image moves the other way.  If we pitch the mirror on the $y$-axis by angle $Y$, the Earth and the image will diverge by angle $2Y$.  But if we roll the mirror on the $x$-axis by angle $X$, the divergence depends on the altitude $\gamma$ of the Sun and Earth.  This is because their projections on the $yz$ plane can be closer to the spin axis.  This is demonstrated in Figure~\ref{fig:RotationTolerance}.  If the Sun and the Earth are on the horizon ($\phi = 180^{\circ}$), rolling the mirror has no effect whatsoever on the alignment between the image and the Earth.  Thus the tolerance to tilting the mirror out of alignment depends on the phase angle and which way it is tilted.\footnote{Yawing the mirror by rotating it around the $z$-axis does not tilt the mirror in and of itself, so it does not change the alignment between the image and the Earth.}

It's important to distinguish misalignment caused by the mirror being tilted and misalignment caused by the relative motion of the Earth and the Sun on the mirror's sky.  Tilting the mirror is what limits $\tStar$, the duration of an individual flash.  Relative motion of Earth and the Sun is what limits $\talign$, the total duration the mirror appears to emit flashes towards Earth (typically a few hours).  This latter quantity is not determined by the mirror's rotation; it's simply the time it takes the Earth to cross the Sun's disk or track on the sky (Figure~\ref{fig:PassageGeometry}).  The difference is that when the mirror itself is rotated, the Earth and the Sun's image necessarily move in opposite directions on the mirror's sky and their relative separation depends on their altitude.  

I now provide a calculation of the size of $\Omega_z$, the solid angle for which the mirror is oriented to display a glint to the Earth.  Again, we can consider by supposing that the mirror starts out perfectly aligned, with the Sun's center and Earth both at altitude $\gamma$, the Sun's center in the $+x$ direction and the Earth in the $-x$ direction.  Thus the Sun's position on the celestial sphere is given by the vector $\hat{r}_{\odot} = (\xsun, \ysun, \zsun) = (\cos \gamma, 0, \sin \gamma)$, and the Earth's vector is $\hat{r}_{\oplus} = (\xearth, \yearth, \zearth) = (-\cos \gamma, 0, \sin \gamma)$.  

Any tilting of the mirror can be accomplished by first yawing the mirror, rotating it on the $z$-axis by angle $Z$, and then pitching on the new $y$-axis by $Y$.  $Y$ can take values between $0$ and $\pi$, while $Z$ can be anything in the range $0$ to $2\pi$.  This way, pure pitch (as described above) happens when $\sin Z = 0$ and pure roll occurs when $\cos Z = 0$.  The rotation matrix is:
\begin{equation}
{\cal Q} = \begin{pmatrix}
                \cos Y \cos Z  & -\cos Y \sin Z  & \sin Y\\
								\sin Z         & \cos Z          & 0\\
								-\sin Y \cos Z & \sin Y \sin Z   & \cos Y,
								\end{pmatrix}
\end{equation}
so the Sun's new position is $\hat{r}^{\prime}_{\odot} = (\xsun^{\prime}, \ysun^{\prime}, \zsun^{\prime}) = (\cos \gamma \cos Y \cos Z + \sin \gamma \sin Y, \cos \gamma \sin Z, -\cos \gamma \sin Y \cos Z + \sin \gamma \cos Y)$, and the Earth's new position is $\hat{r}^{\prime}_{\oplus} = (\xearth^{\prime}, \yearth^{\prime}, \zearth^{\prime}) = (-\cos \gamma \cos Y \cos Z + \sin \gamma \sin Y, -\cos \gamma \sin Z, \cos \gamma \sin Y \cos Z + \sin \gamma \cos Y)$.

The mirror image of the Sun is at the same altitude as the Sun but the opposite azimuth.  The $(x,y,z)$ vector for the image of an object at $\hat{r}$ is $\hat{r}_{\star} = \chi \hat{r}$, where the reflection matrix is
\begin{equation}
\label{eqn:ReflectionMatrix}
\chi = \begin{pmatrix}
                 -1 & 0  & 0\\
								  0 & -1 & 0\\
									0 &  0 & 1
					 \end{pmatrix} .
\end{equation}
Hence, after tilting the mirror, the Sun's beam is aimed at $\hat{r}^{\prime}_{\star} = (-\cos \gamma \cos Y \cos Z - \sin \gamma \sin Y, -\cos \gamma \sin Z, -\cos \gamma \sin Y \cos Z + \sin \gamma \cos Y)$.

A glint is seen if the angle between the new image center and the Earth is smaller than $\thetasun$, with $\hat{r}^{\prime}_{\star} \cdot \hat{r}^{\prime}_{\oplus} \ge \cos \thetasun$.  This condition can be expressed as:
\begin{equation}
\label{eqn:AimEquation}
\cos^2 Y \ge 1 - \frac{\sin^2 (\thetasun/2)}{1 - \cos^2 \gamma \sin^2 Z}.
\end{equation}
For a pure pitching rotation, this means that either $Y \le \thetasun / 2$ or $Y \ge \pi - \thetasun / 2$.  A pure rolling rotation typically allows for a greater freedom in allowed $Y$: $\sin(\thetasun/2)/\sin \gamma \ge |\sin Y|$.  Note that when the Sun and the Earth have $|\gamma| \le \thetasun / 2$, no amount of roll can cause misalignment, as discussed above.

So, for a given phase angle $\phi = \pi - 2\gamma$, the mirror normal can be pointed over a solid angle 
\begin{equation}
\label{eqn:Omegaz}
\Omega_z = 2 \int_0^{2\pi} \left[1 - \sqrt{\max\left(0, 1 - \frac{1 - \cos \thetasun}{1 + \cos^2 Z + \cos \phi \sin^2 Z}\right)}\right] dZ,
\end{equation}
with the $2$ in front standing for both sides of the mirror.  For values of $\phi$ not within a few $\thetasun$ of $\pi$, a useful approximation is $\Omega_z \approx \pi \thetasun^2 / [2 \cos (\phi/2)]$.  

Averaging over all relative orientations of Sun and Earth, I find:
\begin{equation}
\mean{\Omega_z} = \frac{1}{2} \int_{-1}^1 \Omega_z (\phi) d\cos \phi = \pi \thetasun^2,
\end{equation}
in accordance with the symmetry argument at the start of the section.  Note that this $\mean{\Omega_z}$ refers to a probability of a glint occurring, not the amount of light from the glint.  An observer surrounded by a close, isotropic cloud of randomly oriented mirrors would not see a bright region near the Sun  because the higher $\Omega_z$ is countered by the smaller projected area $A_{\rm proj} = A \sin \gamma$.  In fact, the intensity in reflected light from all directions not coincident with the Sun would be the same.  This is a consequence of the fact that, averaged over all orientations and absent diffraction, a reflecting convex solid has the same scattering properties as a sphere with the same surface properties -- which means isotropic scattering for a specularly reflecting shape \citep{vanDeHulst81}.

\section{Calculation of Image Sky Position for a Rotating Mirror}
\label{sec:ImagePath}
Rotation complicates the glint's geometry considerably, since the surface need not be normal to the axis of rotation.  As described in Section~\ref{sec:SunPath} and shown in Figure~\ref{fig:MirrorSkyGeometry}, we can make an analogy to the Earth's celestial sphere, and then also set $\alphasun = 0$ instantaneously.  Then we can define a Cartesian coordinate system with $\hat{x}$ being ``east'' (the direction the Sun rises from), $\hat{y}$ being ``north'' (the pole left of $\hat{u}$), and $\hat{z}$ being the zenith direction.  (Note that these $\hat{x}$ and $\hat{y}$ are not the same as the $\hat{x}$ and $\hat{y}$ of Appendix~\ref{sec:Aim}.)

We can also define an external, inertial coordinate system centered on the surface.  Let $\hat{w}$ be the direction along the rotation axis.  The Sun will be defined to be on the $uw$ plane (with $v_{\odot} = 0$).  In this coordinate system, a spot located at $(\alpha, \delta)$ on the celestial sphere has $u = \cos \alpha \cos \delta$, $v = \sin \alpha \cos \delta$, and $w = \sin \delta$.  So the Sun's direction on the celestial sphere has a vector of $\rVecsun = \cos \decsun \hat{u} + \sin \decsun \hat{w}$.

\begin{figure*}
\centerline{\includegraphics[width=9cm]{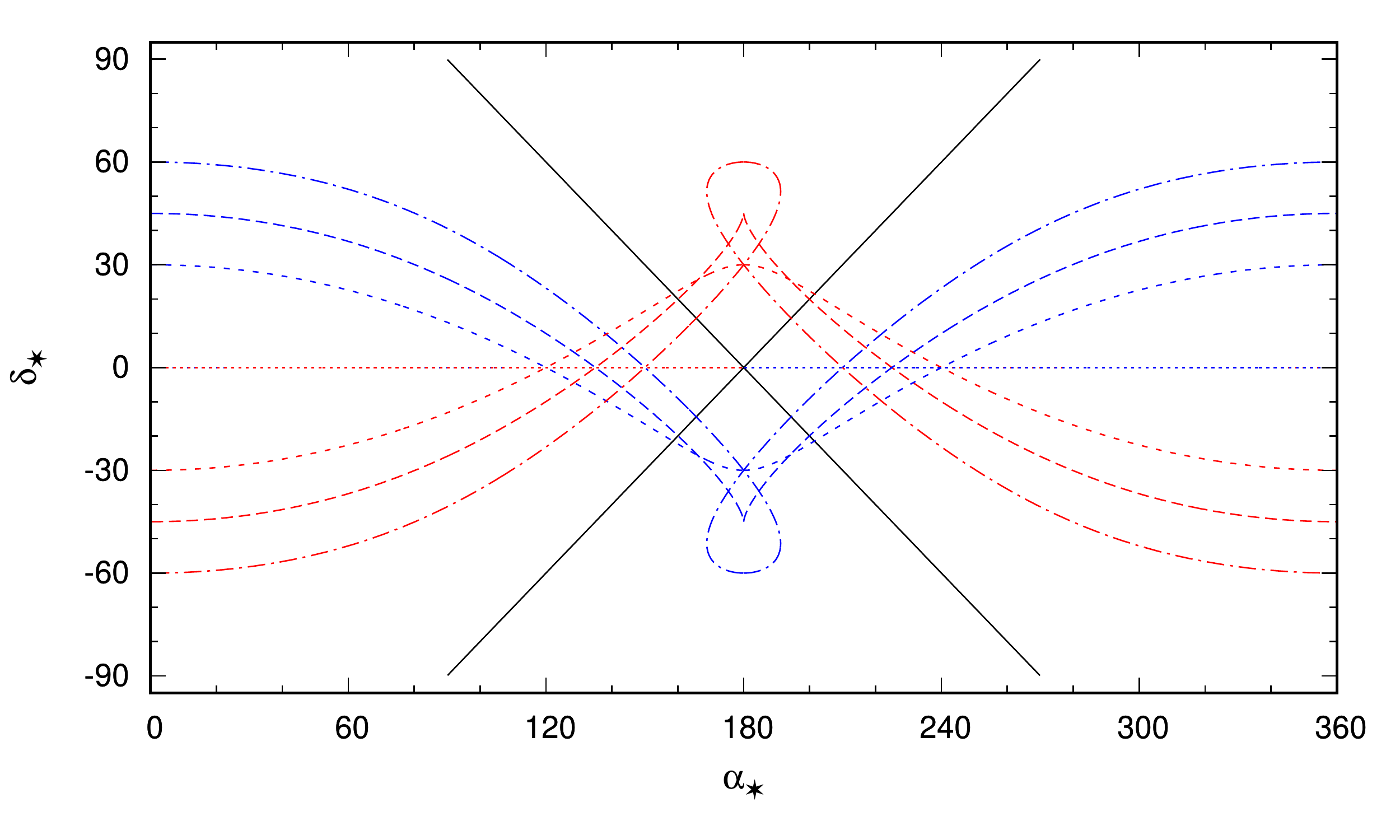}\includegraphics[width=9cm]{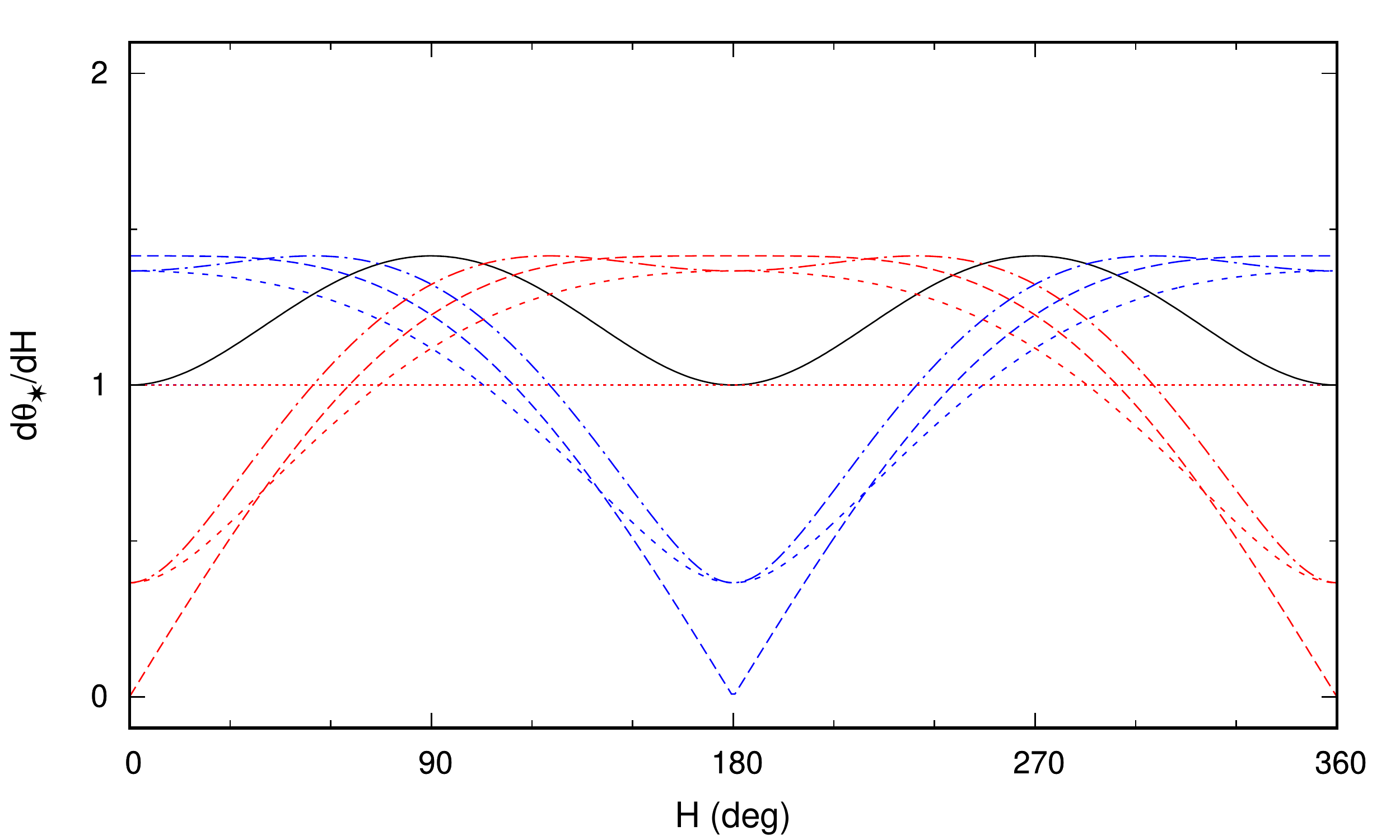}}
\figcaption{\emph{Left}: The path on the sky of the Sun's reflection, as viewed from a spinning artifact with $\Lambda = 45^{\circ}$. \emph{Right:} The angular speed of the Sun's image for the same artifact, in terms of hour angle $H$.  Different Solar declinations have different line styles, with $|\deltasun|$ of $30^{\circ}$ (dash-dot), $45^{\circ}$ (long dashes), $60^{\circ}$ (short dashes), and $90^{\circ}$ (dotted).  Positive declinations are in blue, negative declinations are in red, and an equatorial Sun is denoted by the black solid line.  \label{fig:GlintTrajectories}}
\end{figure*}

The reflection matrix $\chi$ (equation~\ref{eqn:ReflectionMatrix}) allows us to calculate the altitude and azimuth of the image from the Sun's altitude and azimuth.  The direction of the specular reflection in the external, inertial frame can be found by rotating to the comoving frame, reflecting, and then rotating back: $\hat{r}_{\star} = {\cal M} \hat{r}$, where ${\cal M} = {\cal R}^{-1} \chi {\cal R}$.  The transformation from $uvw$ to $xyz$ is found by multiplying by
\begin{equation}
{\cal R} = \begin{pmatrix}
                 -\sin H              & \cos H               & 0\\
								 -\cos H \sin \Lambda & -\sin H \sin \Lambda & \cos \Lambda\\
								  \cos H \cos \Lambda & \sin H \cos \Lambda  & \sin \Lambda
					 \end{pmatrix} .
\end{equation}
So the transformation matrix from $\hat{r}$ to $\hat{r}_{\star}$ is given by:
\begin{equation}
{\cal M}  = \begin{pmatrix}
                2 \cos^2 H \cos^2 \Lambda - 1  & \sin 2H \cos^2 \Lambda        & \cos H \sin 2\Lambda\\
								\sin 2H \cos^2 \Lambda         & 2 \sin^2 H \cos^2 \Lambda - 1 & \sin H \sin 2\Lambda\\
								\cos H \sin 2\Lambda           & \sin H \sin 2\Lambda          & -\cos 2 \Lambda
								\end{pmatrix} ,
\end{equation}
with $\alphastar = \tan^{-1} (v_{\star}/u_{\star})$ and $\deltastar = \sin^{-1} \wstar$.  Finally, the Sun's reflection points at
\begin{equation}
\label{eqn:uvwStar}
\begin{pmatrix} u_{\star} \\ v_{\star} \\ w_{\star} \end{pmatrix} = \begin{pmatrix}
(2 \cos^2 H \cos^2 \Lambda - 1) \cos \deltasun + \cos H \sin 2\Lambda \sin \deltasun \\
\sin 2H \cos^2 \Lambda \cos \deltasun + \sin H \sin 2\Lambda \sin \deltasun \\
\cos H \sin 2\Lambda \cos \deltasun - \cos 2\Lambda \sin \deltasun \end{pmatrix}. 
\end{equation}
The paths traced out by the sun's reflection on the sky are more complicated than simple circles, and can include retrograde loops (Figure~\ref{fig:GlintTrajectories}).

The angular speed of the image on the sky, $d\thetastar/dH = \sqrt{(d\deltastar/dH)^2 + (d\alphastar/dH)^2 \cos^2 \deltastar}$, is a function of $H$, $\Lambda$, and $\rVecsun$ and can be calculated from
\begin{align}
\frac{d\deltastar}{dH} & = -\sin H \sin 2\Lambda \frac{\cos \deltasun}{\cos \deltastar}\\
\frac{d\alphastar}{dH} & = 1 + \frac{\cos^2 \Lambda \cos \deltasun}{\cos^2 \deltastar} \left[\ustar (\cos 2H + \tan^2 \Lambda) + \vstar \sin 2H\right]
\end{align}
The value of $d\thetastar/dH$ when the image of the Sun intersects the Earth determines the duration of each flash.  It is a periodic but not quite sinusoidal function that is typically $\sim 1$ for mid-latitude objects (Figure~\ref{fig:GlintTrajectories}), with a maximum value of $2$ for ``equatorial'' mirrors with $\deltasun = 0$.  This maximum speed corresponds to rotation being a pure pitch discussed in Appendix~\ref{sec:Aim}.

\section{Derivation of Angle Between Sun's Image and Earth for Rotating Mirror}
\label{sec:ImageEarthAngle}
Equation~\ref{eqn:uvwStar} gives the direction of the Sun's image on the external (non-corotating) sky as the mirror rotates.  Starting from the $uvw$ coordinates of Appendix~\ref{sec:ImagePath}, we can rotate the reference frame so that the Sun is at the ``top'' ($\wsun^{\prime} = 1$) by multiplying by the matrix $\Ssun$:
\begin{equation}
\Ssun = \begin{pmatrix}
                 \sin \deltasun & 0 & -\cos\deltasun \\
								  0             & 1 & 0              \\
								 \cos \deltasun & 0 & \sin \deltasun
					 \end{pmatrix} .
\end{equation}
Coordinates in this reference frame will be denoted by $u^{\prime}$, $v^{\prime}$, and $w^{\prime}$.  

At any given moment the Earth is at mirror-defined right ascension $\alphaearth$ and mirror-defined declination $\deltaearth$.  While we can specify the phase angle $\phi$ between the Earth and the Sun from the Solar System geometry, the position angle of the Earth relative to the Sun depends on the mirror orientation.  Since $\alphaearth$ and $\deltaearth$ depend on the arbitrary mirror orientation, it is more convenient to define $\xiearth$ in the $u^{\prime}v^{\prime}w^{\prime}$ frame, with $\uearth^{\prime} = \cos \xiearth \sin \phi$, $\vearth^{\prime} = \sin \xiearth \sin \phi$, and $\wearth^{\prime} = \cos \phi$.

To calculate the angular distance between the Sun's image and the Earth on the mirror sky, we rotate again from the $u^{\prime}v^{\prime}w^{\prime}$ frame to the $u^{\dprime}v^{\dprime}w^{\dprime}$ frame, where the Earth is on ``top'' with $\wearth^{\dprime} = 1$.  This rotation matrix is given by:
\begin{equation}
\Searth^{\prime} = \begin{pmatrix}
                 \cos \phi \cos \xiearth & \cos \phi \sin \xiearth & -\sin \phi \\
								 -\sin \xiearth          & \cos \xiearth           & 0\\
								 \sin \phi \cos \xiearth & \sin \phi \sin \xiearth & \cos \phi
					 \end{pmatrix} ,
\end{equation}
where $(u^{\dprime}, v^{\dprime}, w^{\dprime})^T = \Searth (u^{\prime}, v^{\prime}, w^{\prime})^T$.  The coordinates of the Sun's image in this system are found from $(\ustar^{\dprime}, \vstar^{\dprime}, \wstar^{\dprime})^T = \Searth^{\prime} \Ssun (\ustar, \vstar, \wstar)^T$:
\begin{equation}
\begin{pmatrix} \ustar^{\dprime} \\ \vstar^{\dprime} \\ \wstar^{\dprime} \end{pmatrix} = \begin{pmatrix}
\cos \xiearth \cos \phi \sin \deltasun - \sin \phi \cos \deltasun & \sin \xiearth \cos \phi & -\cos \xiearth \cos \phi \cos \deltasun - \sin \phi \sin \deltasun\\
-\sin \xiearth \sin \deltasun & \cos \xiearth & \sin \xiearth \cos \deltasun\\
\cos \xiearth \sin \phi \sin \deltasun + \cos \phi \cos \deltasun & \sin \xiearth \sin \phi & -\cos \xiearth \sin \phi \cos \deltasun + \cos \phi \sin \deltasun \end{pmatrix} \begin{pmatrix} \ustar \\ \vstar \\ \wstar \end{pmatrix} .
\end{equation}

The angle between the center of the Sun's image and the Earth at any moment is found from $\cos \theta = \hat{r}_{oplus} \cdot \hat{r}_{\star} = \wstar^{\dprime}$.  Plugging in the image coordinates from equation~\ref{eqn:uvwStar}, I find:
\begin{align}
\label{eqn:wDoublePrime}
\nonumber \wstar^{\dprime} & = \cos^2 H \cos^2 \Lambda [\cos \phi (1 + \cos 2\deltasun) + \cos \xiearth \sin \phi \sin 2\deltasun] + \cos H \sin 2 \Lambda [\cos \phi \sin 2\deltasun - \cos \xiearth \sin \phi \cos 2\deltasun] \\
\nonumber                  & + \sin 2H \cos^2 \Lambda (\sin \xiearth \sin \phi \cos\deltasun) + \sin H \sin 2\Lambda (\sin \xiearth \sin \phi \sin \deltasun) \\
													 & + [\cos \phi (2 \sin^2 \Lambda \sin^2 \deltasun - 1) - \sin^2 \Lambda \cos \xiearth \sin \phi \sin 2\deltasun] .
\end{align}
A glint is seen when $\wstar^{\dprime} \ge \cos \thetasun$.  Equation~\ref{eqn:wDoublePrime} is a quartic equation for general values of $\phi$ and $\xiearth$.  The solution of $\wstar^{\dprime} = \cos \thetasun$ is therefore extremely complicated.  At opposition ($\phi = 0$), however, it reduces to a quadratic equation in $\cos H$, and $\wstar^{\dprime} = \cos \thetasun$ happens when $H = H_{\rm edge}$ in the following relation:
\begin{equation}
\cos^2 \Hedge (2 \cos^2 \Lambda \cos^2 \deltasun) + \cos \Hedge (\sin 2\Lambda \sin 2\deltasun) + (2 \sin^2 \deltasun \sin^2\Lambda - 1 - \cos \thetasun) = 0.
\end{equation}
Equation~\ref{eqn:HStarOpposition} follows from this relation by finding the two solutions for $\cos \Hedge$, $\cos \Hedge^-$ (smaller) and $\cos \Hedge^+$ (larger), and then verifying that $w^{\dprime} > \cos \thetasun$ only when $-1 \le \cos H < \cos \Hedge^-$ for $\deltasun < 0$ or $\cos \Hedge^+ < \cos H \le 1$ for $\deltasun > 0$. 

\acknowledgements
I wish to acknowledge and thank the Breakthrough Listen program for their support.  Funding for \emph{Breakthrough Listen} research is sponsored by the Breakthrough Prize Foundation\footnote{https://breakthroughprize.org/}.

In addition, I acknowledge this work's use of NASA's Astrophysics Data System and arXiv.

%========

\end{document}